\documentclass[12pt,iop,onecolumn]{emulateapj}
\pdfoutput=1
\usepackage{apjfonts}

\newcommand{\src}{G11.2$-$0.3}
\catcode`\@=11
\newcommand{\gapprox}{\mathrel{\mathpalette\@versim>}}
\newcommand{\lapprox}{\mathrel{\mathpalette\@versim<}}
\newcommand{\propapprox}{\mathrel{\mathpalette\@versim\propto}}
\newcommand{\@versim}[2]
  {\lower3.1truept\vbox{\baselineskip0pt\lineskip0.5truept
\ialign{$\m@th#1\hfil##\hfil$\crcr#2\crcr\sim\crcr}}}
\catcode`\@=12

\shorttitle{Young Core-Collapse SNR G11.2$-$0.3}

\begin{document}

\title{G11.2$-$0.3: The Young Remnant of a Stripped-Envelope Supernova}

\author{Kazimierz J. Borkowski,\altaffilmark{1}
Stephen P. Reynolds,\altaffilmark{1}
\&\ Mallory S. E. Roberts \altaffilmark{2}
}

\altaffiltext{1}{Department of Physics, North Carolina State University, 
Raleigh, NC 27695-8202; kborkow@unity.ncsu.edu} 
\altaffiltext{2} {New York University Abu Dhabi, U.A.E.}

\submitted{Accepted for publication in ApJ}

\begin{abstract}
We present results of a 400-ks {\sl Chandra} observation of the young
shell supernova remnant (SNR) \src, containing a pulsar and
pulsar-wind nebula (PWN).  We measure a mean expansion rate for the
shell since 2000 of $0.0277 \pm 0.0018$\%\ yr$^{-1}$, implying an age between
1400 and 2400 yr, and making \src\ one of the youngest core-collapse
SNRs in the Galaxy.  However, we find very high absorption 
($A_V \sim 16^m \pm 2^m$),
confirming near-IR determinations and ruling out a claimed
association with the possible historical SN of 386 CE.  The PWN shows
strong jets and a faint torus within a larger,
more diffuse region of radio emission and nonthermal X-rays. Central
soft thermal X-ray emission is anticorrelated with the PWN; that, and
more detailed morphological evidence, indicates that the reverse shock
has already reheated all ejecta and compressed the PWN.  The pulsar
characteristic energy-loss timescale is well in excess of the remnant
age, and we suggest that the bright jets have been produced since the
recompression.  The relatively pronounced shell and diffuse hard X-ray
emission in the interior, enhanced
at the inner edge of the shell, 
indicate that the immediate circumstellar medium into
which \src\ is expanding was quite anisotropic.  We propose a possible
origin for \src\ in a stripped-envelope progenitor that had lost almost all
its envelope mass, in an anisotropic wind or due to binary
interaction, leaving a compact core whose fast winds swept previously
lost mass into a dense irregular shell, and which exploded as a Type
cIIb or Ibc supernova. 

\end{abstract}


\section{Introduction}
\label{intro}

The remnants of core-collapse supernovae (CC SNRs) contain a wealth of
information on the progenitor system, the supernova (SN) event, and
the surrounding material.  Young CC SNRs show the clearest imprint of
the SN itself and the composition of the ejecta.  If a neutron star
left behind acts as a pulsar, it can inflate a pulsar-wind nebula
(PWN) in the midst of the expanding ejecta.  Young PWNe are of great
interest in their own right, as they exhibit the particles and
magnetic field generated in the relativistic pulsar outflow and
modified in a relativistic wind termination shock.  However, the PWN
can also act as a probe of the innermost material ejected in the
supernova.

The youngest CC SNR in the Galaxy, Cas A, contains a neutron star
which unfortunately does not function as a pulsar, while the youngest
PWN in the Galaxy, the Crab Nebula, shows only very indirect evidence
for the surrounding SNR shell that ought to contain it (see Hester
2008 for a review).  The other historical SNRs, Kepler, Tycho, SN
1006, RCW 86 (= SN 185 CE; Williams et al. 2011) and G1.9+0.3 (the
latter only quasi-historical, as the SN event was not observed at
Earth due to obscuration around 1900 when it would have been seen),
all appear to be remnants of thermonuclear, Type Ia events.  For years
it was thought that the next youngest historical CC SNR was a bright
shell radio source, G11.2$-$0.3; in fact, one of the present authors
\citep{reynolds94} used ROSAT observations to claim support for its
association with a possible supernova in CE 386 (Stephenson \& Green
2002), as well as arguing for a Type Ia origin.  Both these assertions
have turned out to be incorrect.

G11.2$-$0.3 has been identified as a potential historical SNR since at
least the first well-resolved imaging observations at radio (VLA,
$20''$ resolution) and X-ray ({\sl Einstein} HRI) wavelengths
\citep{downes84}, where it was argued that its high radio surface
brightness and symmetrical well-defined shell structure indicated
youth.  An identification with the CE 386 event was proposed, but an
even younger age was also suggested.  Green et al.~(1988) observed
\src\ with the VLA with $3''$ resolution at two frequencies, and
argued on morphological grounds that \src\ resembled Cas A much more
than Tycho or Kepler, in particular in the clumpy structure and lack
of a sharp outer boundary.  They reinterpreted an H I absorption
spectrum of \cite{becker85} to estimate a distance of about 5 kpc, and
also supported an association with the CE 386 event. More recent distance
estimates range from 4.4 kpc \citep{green04} to 5.5--7 kpc \citep{minter08}.
(Below, we use a distance of 5 kpc but also consider variations 
between 4.4 and 7 kpc.)

The question of the SN type giving rise to \src\ was definitively
settled with the discovery with {\sl ASCA} of the PWN
\citep{vasisht96} and the 65 ms pulsar \citep{torii97}.  However, the
measured spindown \citep{torii99} gave a characteristic age $\tau_c
\equiv P/2{\dot P}$ of about 20,000 yr, implying that the pulsar was
born spinning at nearly its present period (so $\tau_c$ is a poor age
indicator).  Best values for the pulsar period and period derivative
were obtained with RXTE (see the {\sl Fermi} Ephemerides 
database\footnote{
http://fermi.gsfc.nasa.gov/ssc/data/access/lat/ephems/index.html}):
$P = 64.69$ ms and $\dot
P = 4.43 \times 10^{-14}$ s s$^{-1}$, giving $\tau_c = 23,000$ yr and
spindown luminosity $\dot E = 6.5 \times 10^{36}$ erg s$^{-1}$.

Further radio spectral studies \citep{kothes01} confirmed the presence
of a flat-spectrum radio core ($\alpha \sim 0$ with $S_\nu \propto
\nu^{\alpha}$) with observations between 4.85 and 32 GHz, using the
Effelsberg 100-m telescope.  The most detailed radio observations of
\src\ were those of \citet{tam02} and \citet{tam03} with the VLA.
\citet{tam02} reported a PWN spectral index of
$\alpha = -0.25^{+0.05}_{-0.10}$
and a shell spectrum with $\alpha = -0.56 \pm 0.02$, from
archival VLA observations.  The highest resolution imaging of the PWN
radio structure was reported in \citet{roberts03}.
The radio image
shows loops and arcs of radio emission most prominent to the
northeast (Figures~\ref{pwnextent}, \ref{pwncloseup}, and \ref{rxtori}).
While the pulsar
dominates the X-ray image, it is not detectable in the radio image.
This is consistent with the upper limit of 0.1 mJy on the pulsed flux
obtained from deep 1.9 GHz radio searches with the
Green Bank Telescope (H.~Al Noori, private communication). 

\src\ was an early target of the {\sl Chandra} X-ray Observatory.
\citet{kaspi01} showed that the pulsar was within $8''$ of the
geometric center of the remnant, providing additional evidence that
the pulsar spindown age greatly overestimated the true age of the
pulsar and remnant.  \cite{roberts03} analyzed the morphology and
spectrum of the shell and PWN, in comparison with high-resolution
radio data.  They focused primarily on the PWN, showing the radio
counterpart, and pointing out small-scale changes in X-ray morphology
between observations in August and October 2000.  Spectral analysis
showed that a plane-shock model
did a reasonable job
of describing the brightest (SE) portion of the shell, with $kT \sim
0.6$ keV and an ionization timescale of about $7 \times 10^{11}$
cm$^{-3}$ s.  However, a hard excess was present, which they described
with a model of synchrotron emission from a power-law electron
spectrum with an exponential cutoff,
requiring a rolloff
frequency (the characteristic synchrotron frequency emitted by
electrons with the e-folding energy $E_{\rm max}$ of the exponential
cutoff) of about $1.8 \times 10^{16}$ Hz.  A further {\sl Chandra}
observation of 60 ks was obtained in 2003, confirming variability of the
PWN \citep{roberts04}. X-ray emission from the PWN was found to consist of
bright jets and much fainter, more diffuse emission mostly filling the
radio PWN \citep{roberts05}.

\src\ was reported as an infrared source in the IRAS catalogue of
\citet{arendt89}, but this was probably unrelated emission from an H
II region \citep{reach06}.  \citet{reach06} detected \src\ with {\sl
  Spitzer} in the GLIMPSE survey of the Galactic plane with the IRAC
instrument.  Only faint filamentary emission in a few spots was
detected, primarily at the longer wavelengths (5.8 and 8 $\mu$m).
However, at 24 $\mu$m, {\sl Spitzer} MIPS observations show a bright,
complete shell corresponding well to the radio and X-ray structure and
attributed to emission from collisionally heated grains
\citep{pinheiro11,andersen11}.  Recent near-IR observations (Koo et
al.~2007; Moon et al.~2009) show [Fe II] emission at 1.644 $\mu$m from
the bright SE shell and the fainter NW shell as well as in several
knots surrounding the PWN.  The outer shell emission may represent a
mixture of ejecta and shocked circumstellar medium (CSM).  The
interior emission shows Doppler shifts of up to 1000 km s$^{-1}$, and
is interpreted as inner ejecta (Moon et al.~2009) in undecelerated
expansion. More spatially-localized emission from molecular hydrogen
is also present \citep{koo07,froebrich15}, with several H$_2$
filaments located outside the radio boundary of the remnant. The recent
detection of broad CO lines provides additional evidence for the presence of
shocked molecular gas in \src\ \citep{kilpatrick16}.

G11.2$-$0.3 shows expansion at radio and near-IR wavelengths.  Tam \&
Roberts (2003) derived radio expansion rates of $0.057'' \pm 0.012''$
yr$^{-1}$ and $0.040'' \pm 0.013''$ yr$^{-1}$ at 20 and 6 cm,
respectively, while Koo et al.~(2007) found the SE [Fe II] filament to
be expanding at $0.035'' \pm 0.013''$ yr$^{-1}$.  While the errors are
large, these rates imply undecelerated ages (that is, upper limits) of
4,000 -- 7,000 yr, confirming the relative youth of \src.

\src\ presents various interesting problems for the evolution of SNRs
and their massive-star progenitors.  \citet{chevalier05} included it
in a group of remnants of Type IIL/b supernovae, a class of supernovae
which have lost most of their hydrogen envelopes at the time of
explosion, implying pre-explosion mass loss of at least several
$M_\odot$.  He interpreted the observations as supporting a SN IIL/b
event in which the shell's reverse shock has not yet reached the PWN.
If \src\ originated in such a supernova, the current SNR blast wave
should be encountering modified CSM, perhaps in
a steady-state wind with $\rho \propto r^{-2}$.  The large extent of
the PWN relative to the shell has been used as evidence that the
reverse shock has not quite reached the PWN yet \citep{tam02}, but the
nature of central thermal X-ray emission
\citep[see Figure 1 in][]{roberts03} 
is not
at all clear.  So even basic questions concerning this important
remnant are not definitively answered: What is the actual age?  Is it
encountering stellar-wind material, undisturbed interstellar medium (ISM), 
or something
else?  Where is the reverse shock?  Is there evidence in X-rays for
ejecta emission, whose abundances might contain clues to the
progenitor mass?  We address these questions below with a $\sim 400$ ks {\sl
  Chandra} observation.

\section{X-Ray Observations}
\label{obsxrays}

{\it Chandra} deep observations of \object{G11.2$-$0.3} took place in 
2013 May and September (Table \ref{observationlog}), with the remnant 
located on the ACIS S3 CCD chip.  All 
data have been reprocessed with CIAO v4.6 and CALDB v4.6.3, and screened for 
periods of high particle background. Very Faint mode was used as the 
surface brightness of \object{G11.2$-$0.3} is low, allowing for an 
efficient rejection of particle background. The total effective exposure time 
is 388 ks ($3/4$ in May and $1/4$ in September). Observations in May consist of 
one long (173 ks) pointing in its first week, and two shorter pointings
(combined exposure of 121 ks) in its last week. This makes it possible to study 
the short-term variability of the PWN. Earlier observations from 2000 and 
2003 (Table \ref{observationlog}) have also been used for investigation of the 
long-term variability of the PWN and for measuring the remnant's expansion. The 
same processing steps were followed as for the 2013 observations. The earliest 
observation from 2000 August was done in Faint mode, so its particle background 
is higher than for other datasets. 

\begin{deluxetable}{lccc}
\tablecolumns{4}
\tablewidth{0pc}
\tablecaption{{\sl Chandra} Observations of G11.2$-$0.3}
\tablehead{
\colhead{Date} & Observation ID & Roll Angle & Effective exposure time \\
& & (deg) & (ks) }

\startdata
2000 Aug 06  & 780 & 267  & 19 \\
2000 Oct 15 & 781 & 273 & 10 \\
2000 Oct 15 & 2322 & 273 & 4.6  \\
2003 May 10  & 3909 & 95 & 14 \\
2003 Jun 27 & 3910 & 209 & 14 \\
2003 Aug 01 & 3911 & 258 & 7.3  \\
2003 Sep 08 & 3912 & 270 & 15 \\
2013 May 05--07 & 14831 & 95 & 173  \\
2013 May 25--26 & 14830 & 98 & 58  \\
2013 May 26--27 & 14832 & 98 & 63  \\
2013 Sep 07 & 15652 & 270 & 48  \\
2013 Sep 08--09 & 16323 & 270 & 46  
\enddata
\label{observationlog}
\end{deluxetable}

We aligned the 2013 observations to match the reference frame of the longest 
pointing from early May (Obs.~ID 14831). Most detected X-rays are produced by  
\object{G11.2$-$0.3}, so we smoothed the 2013 May 5--7 data with the 
multiscale partitioning method of \citet{krishnamurthy10}, and then used 
the method described in \S~\ref{expansion} to align the other 2013 pointings. 
The coordinate transformation used involves only simple translations, without 
any rotation or change in the physical scale. Alignment of the 2000 and 2003 
observations to the 2013 reference frame was done simultaneously with the 
measurement of the remnant's expansion (as described in \S~\ref{expansion}). 
This involves change in the physical scale but not rotation.

Spectral analysis was done with XSPEC v12.8.2
\citep{arnaud96}. Background was extracted from a large area on the S3
CCD chip away from \object{G11.2$-$0.3}. The background was modeled
instead of subtracted in order to allow the use of C-statistics
\citep{cash79}.

\section{Radio Observations}
\label{obsradio}

The Very Large Array observed the field containing G11.2$-$0.3 in
2001--2002 at 20~cm, 6~cm, and 3.5~cm. The 20~cm and 6~cm observations
and reduction are described in \citet{tam03} and were used to measure
the expansion of the remnant. Since the PWN has a significantly
flatter spectrum than the shell, it is more prominent at 3.5~cm, and
we use those data for the morphological comparisons described here. The
same image was previously used in \citet{roberts03}.

A four-pointing mosaic centered on the remnant was used with the
pointings separated by half a primary beamwidth. The bandwidth was
100~MHz which, after flagging and averaging the data sets, had an
effective central frequency of 8459.4~MHz. The observations were
interleaved with the observations at 20~cm and 6~cm presented in
\citet{tam03} so as to maximize the hour angle coverage. Each pointing
of the mosaic had exposures of roughly 27~min in the DnC Array, 18~min
in the C array, 13~min in the CnB array, and 31~min in the BnA
array. The data processing was performed using standard procedures
within the {\it MIRIAD} package \citep{sk99}\footnote{See 
http://www.atnf.csiro.au/computing/software/miriad} in mosaic and
multifrequency synthesis mode. We performed calibration and editing on
each data set individually, before combining all the data. The primary
gains were determined using 3C 286 and 3C 48, and phase calibrations
were made from observations of 1820-254 (J2000.0). Imaging was
performed with Robust weighting as a compromise between maximized
signal-to-noise ratio and resolution. We utilized the maximum entropy
method algorithm for deconvolution \citep{cbb99} and applied
self-calibration iteratively to improve phase and amplitude
calibrations. The final image is corrected for primary-beam
attenuation and has a synthesized beam of $3.1\arcsec \times
2.6\arcsec$. This image was then further smoothed with various kernels
to bring out specific features to compare with X-ray images.

\section{Pulsar-Wind Nebula}

\begin{figure}
\includegraphics[angle=0,scale=0.505]{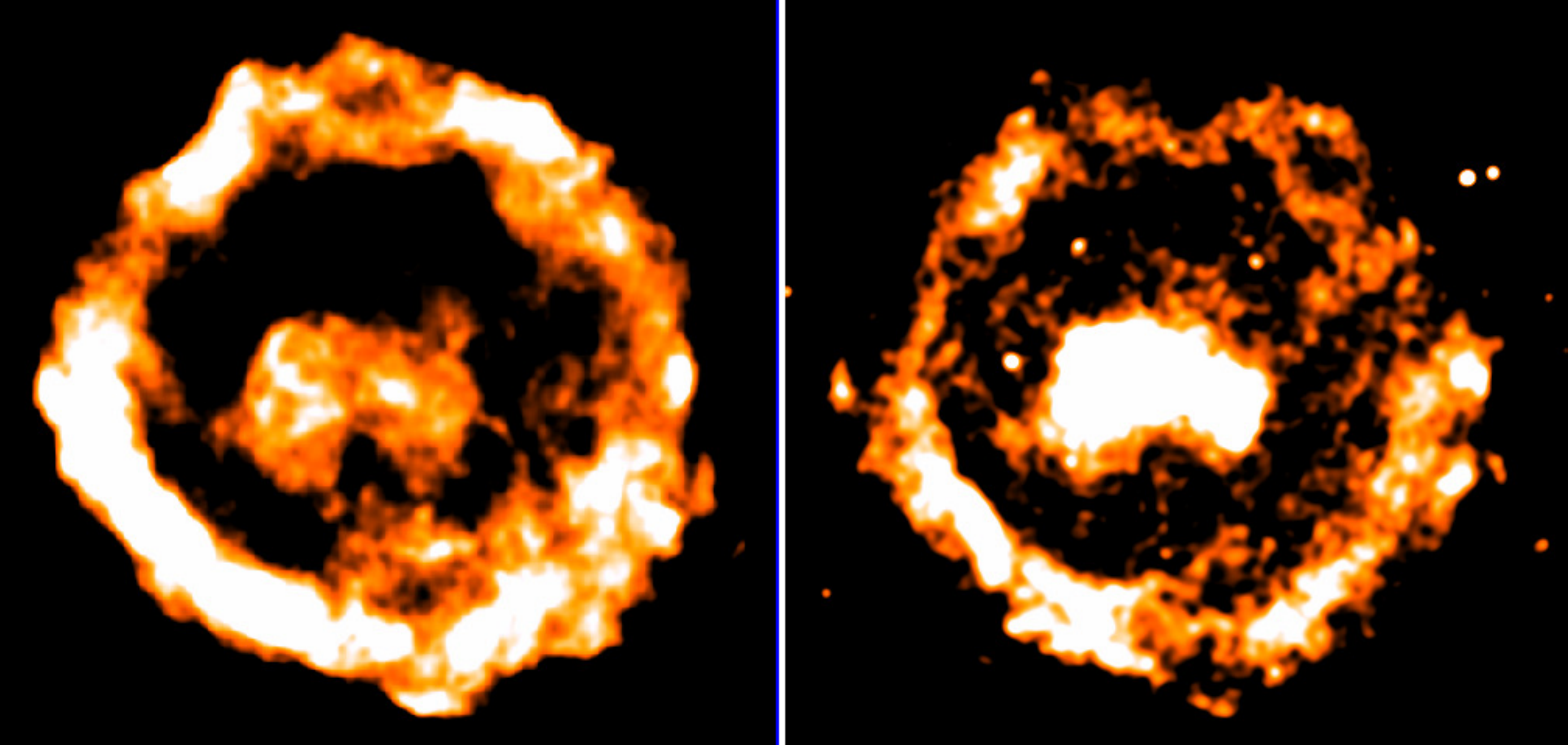}
\caption{Left: Smoothed 3.5 cm VLA image.  Right: Smoothed 3.3 -- 8.1
  keV {\sl Chandra} image.  Both are scaled to show the similar
  maximum extent of the central pulsar-wind nebula.  The pulsar itself is
much brighter and has
been masked out in the X-ray image; it is not visible in the radio image.
N is up and E is to the left. }
\label{pwnextent}
\end{figure}

\begin{figure}
  \center{
  \includegraphics[angle=0,scale=0.7]{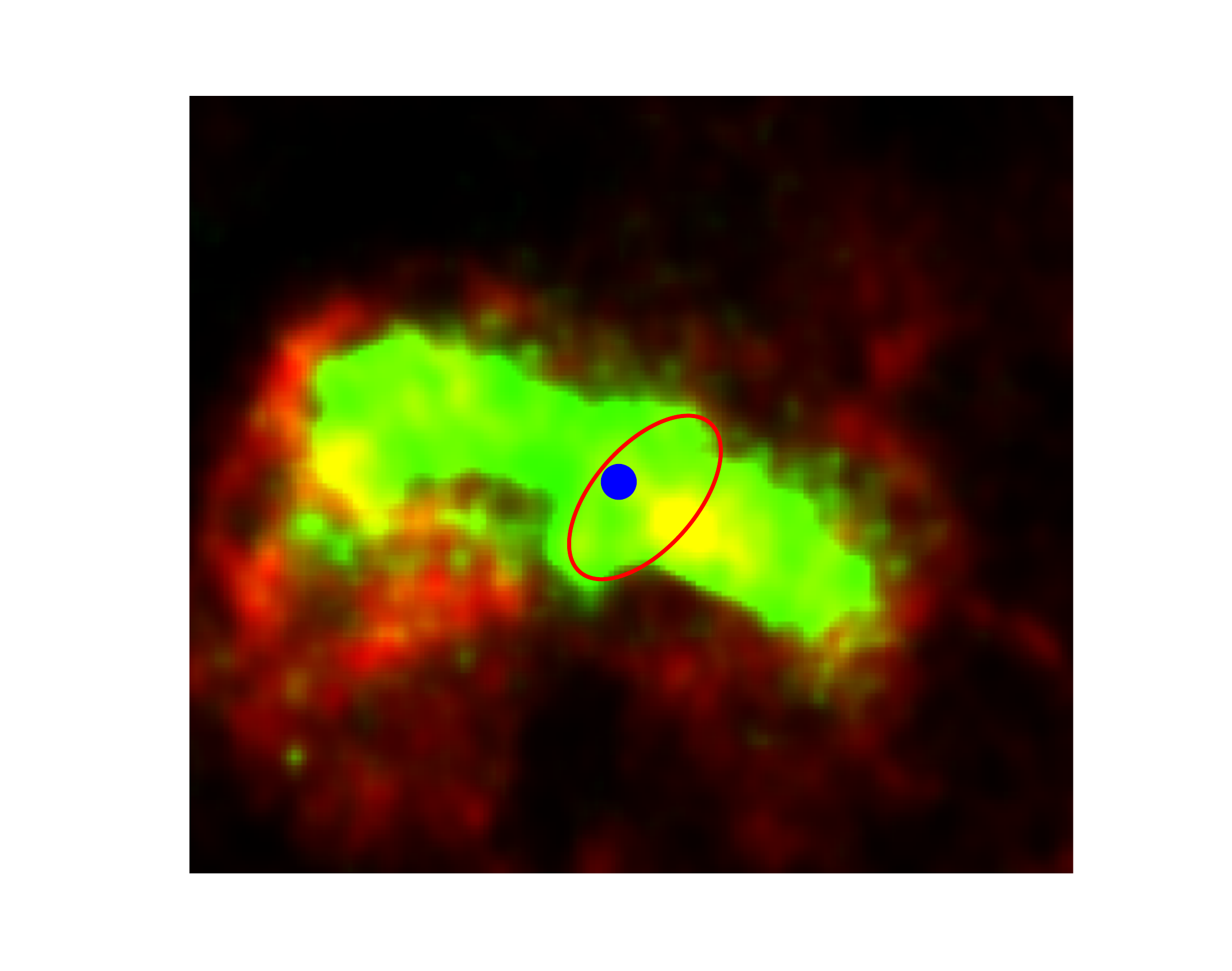}
\caption{Red: 3.5 cm radio image.  Note filamentary arcs and loops,
  especially to the east.  Green: 3.3 -- 8.1 keV image.  The
  morphology is simpler, mainly a two-sided jet.  The pulsar has been
  masked out (blue region); an ellipse fit to the radio torus is
overplotted.}
\label{pwncloseup}
}
\end{figure}

\cite{roberts03} and \cite{tam02} presented the basic properties of
the PWN at radio and X-ray wavelengths. The central radio emission
shown in Fig.~\ref{pwnextent} has a spectral index $\alpha$
of $\alpha_P = -0.25_{-0.10}^{+0.05}$, while the
mean spectral index of the shell is $\alpha_S = -0.56 \pm 0.02$,
although there are spatial variations.  While the extents of the radio
and X-ray PWN are comparable, the interior structures differ
substantially (Fig.~\ref{pwncloseup}). We estimate the length of the
X-ray jets to be about $35''$; they do not have clear radio
counterparts.  The jet widths are of order $10'' - 15''$.  (At 
5 kpc, $1'' = 7.5 \times 10^{16}$ cm.) The radio
emission is mainly in filamentary arcs and loops; the outermost loop
to the NE appears to bound the X-ray jet.

\begin{figure}
{\hskip-0.5truein\includegraphics[angle=0,scale=0.55]{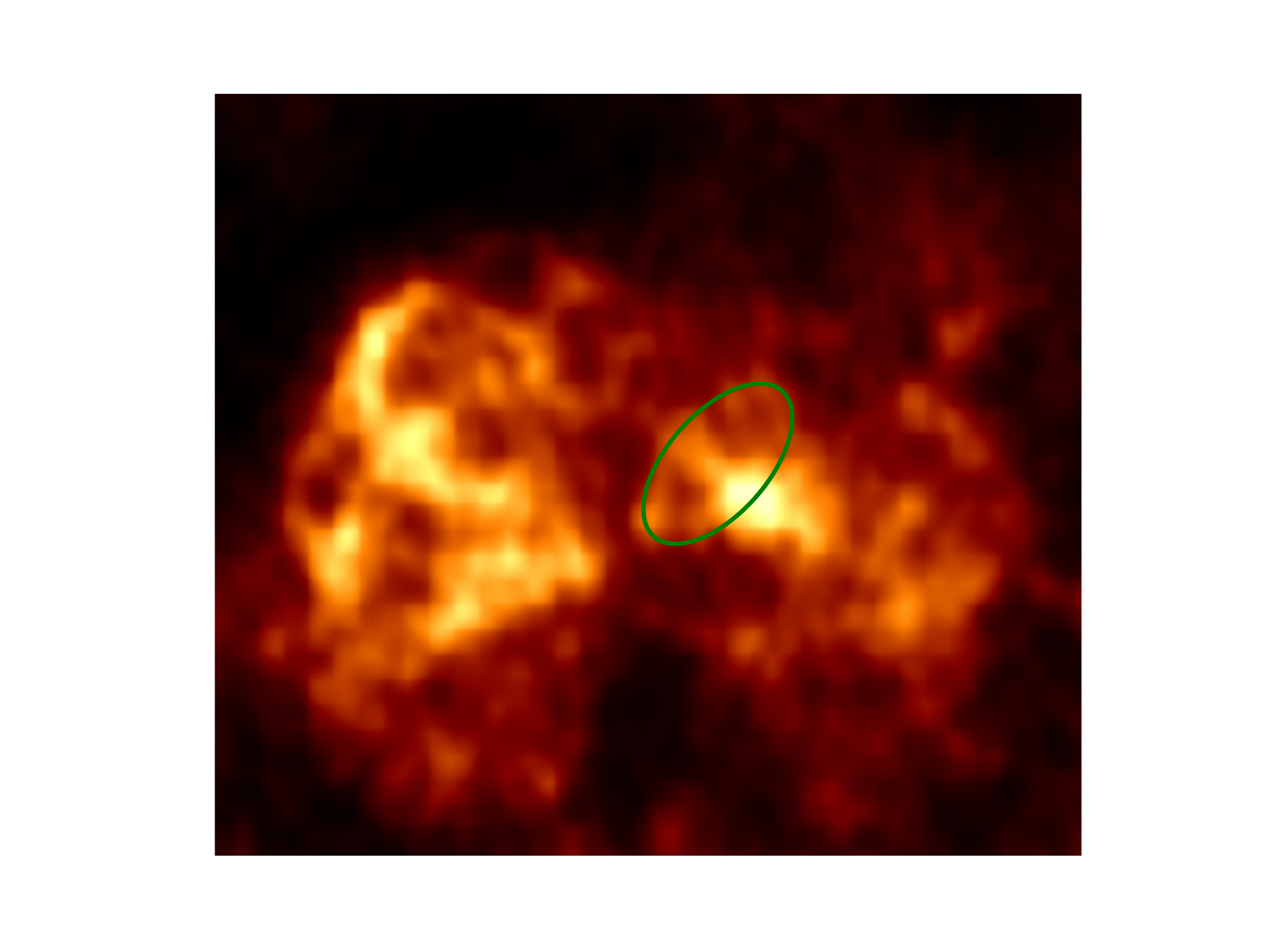} 
  \hskip-0.9truein \includegraphics[angle=0,scale=0.55]{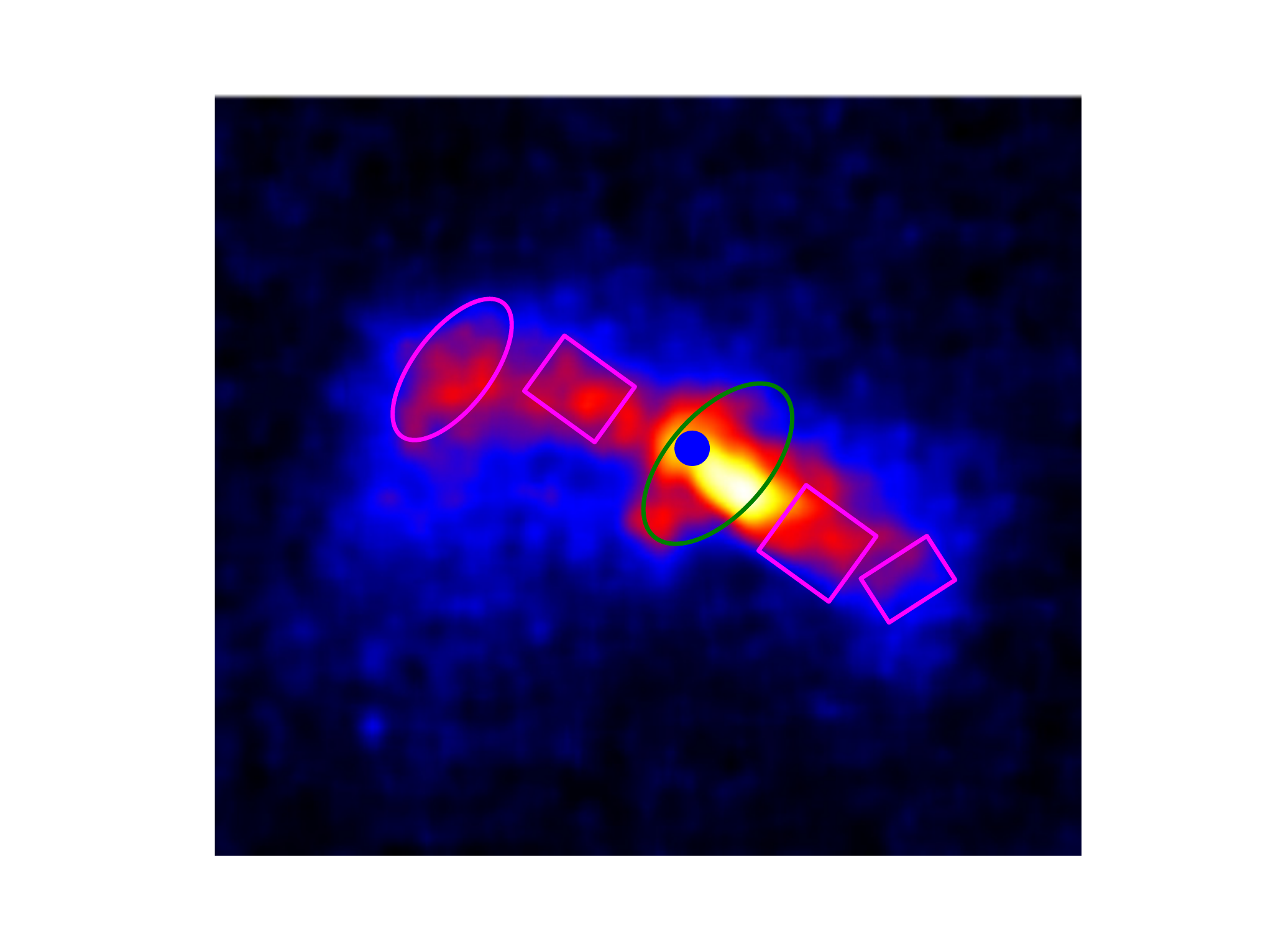} }
\caption{Left: 3.5 cm radio image with torus indicated.  Right:
  Smoothed {\sl Chandra} image from 3.3 to 8 keV, with radio torus
  indicated in green, on arcsinh stretch scale.  Note the structure 
  coincident with
  the radio torus.  The radio maximum is not the pulsar but the
  intersection of the torus with the SW jet; the X-ray pulsar has
  again been masked out. Spectral extraction regions along the jets are also
  shown.}
\label{rxtori}
\end{figure}

\begin{figure}
  \center{
  \includegraphics[angle=270,scale=0.5]{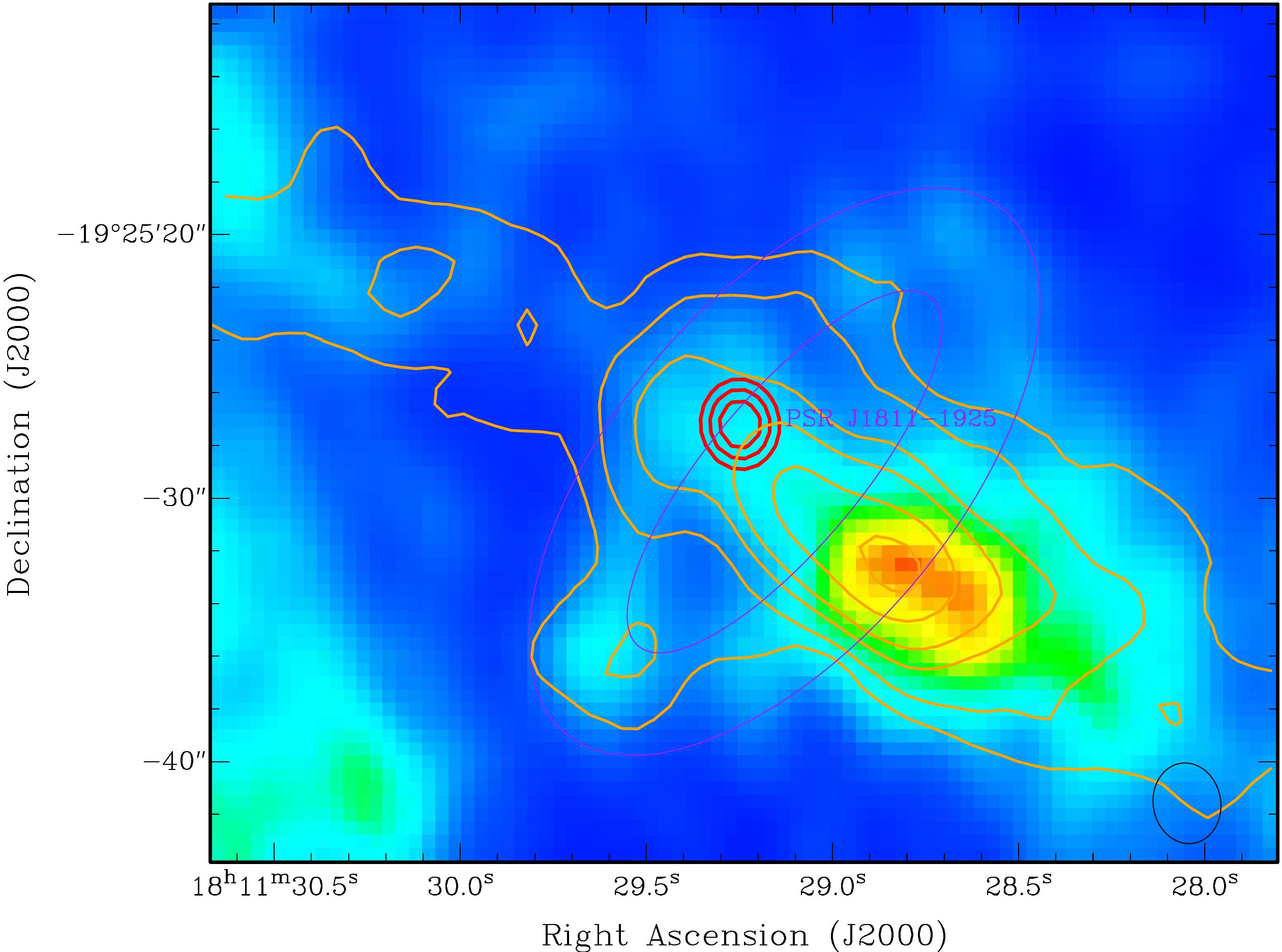}
  \caption{A close-up of the torus/SW jet region in the 3.5~cm image with colors
chosen to highlight the torus and bright spot. The orange contours are
from the combined 3.3--8 keV Chandra  image. Note that the contours are
not equally spaced in intensity, but chosen to highlight particular
structures in the X-ray PWN  which correspond to structures in the
radio without unduly obscuring the radio image. The purple ovals
outline the proposed radio torus, while the red contours show the
location of the pulsar  in X-rays. The synthesized radio beam is
indicated by the black oval in the bottom right corner.}
  \label{radiotorus}
  }
  \end{figure}

\begin{figure}
\includegraphics[angle=0,scale=0.29]{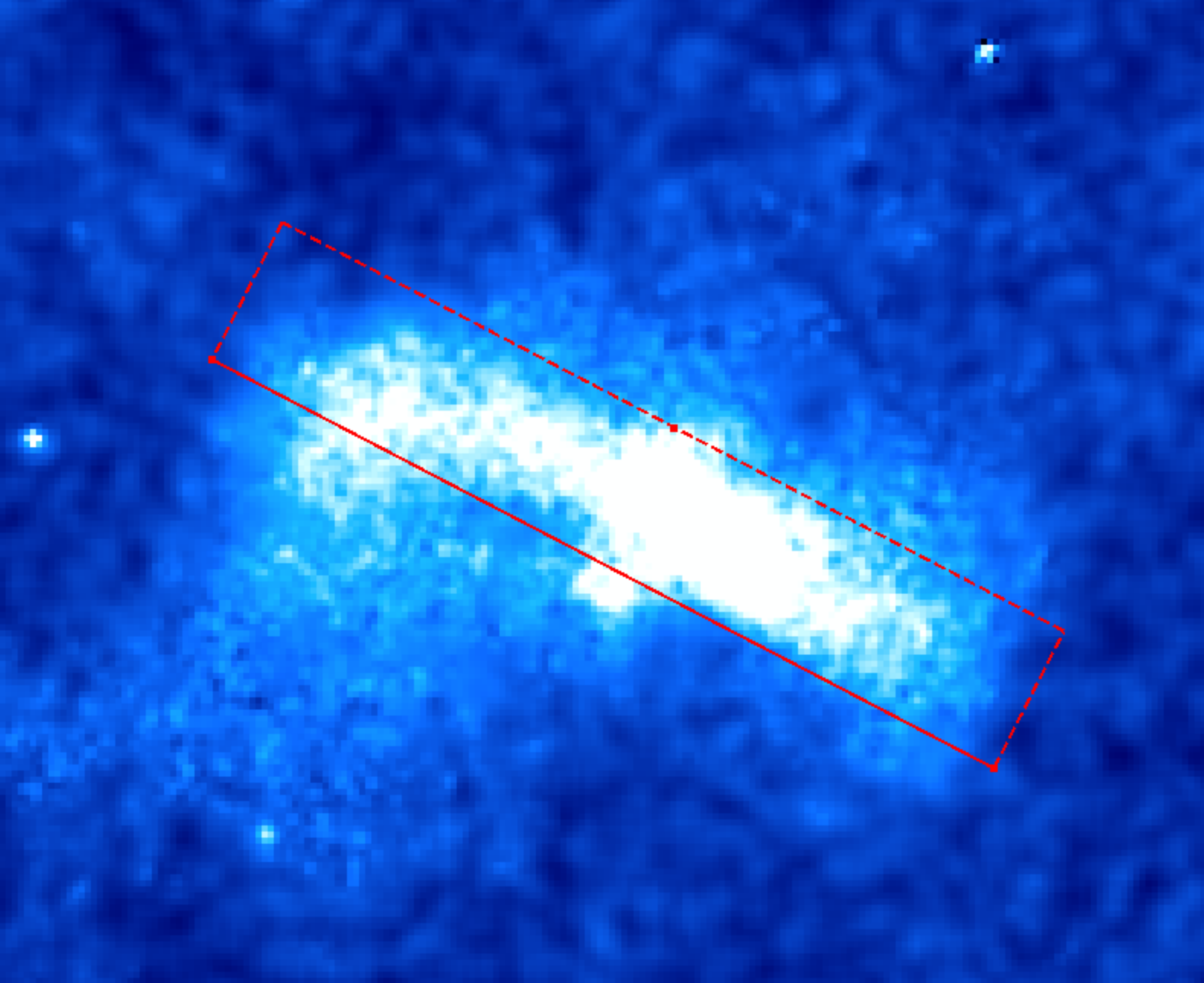}\hskip0.1truein
\includegraphics[angle=0,scale=0.54]{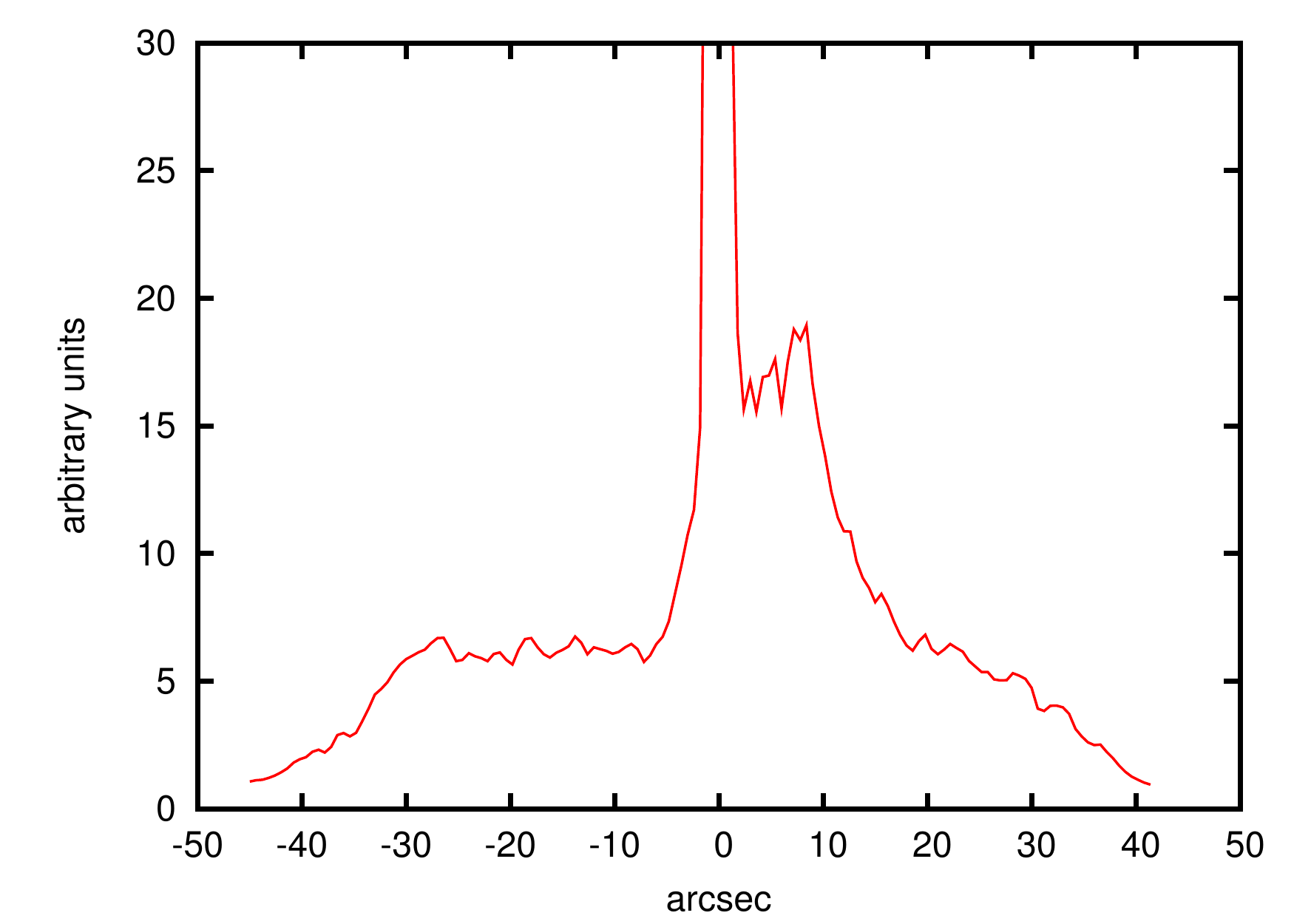}
\caption{Left: Region used for jet profile (summed along the short
  direction).  The region is $87'' \times 15''$, at a position angle
of $62^\circ$ E of N.  
 Right: X-ray jet profile from NE to SW through 3.3 -- 8
  keV image (pulsar not removed).  The bright spots to the SW are
  about three times as bright as the mean emission to the NE.}
\label{jetprofile}
\end{figure}

\begin{figure}
  \center{
  \includegraphics[angle=270,scale=0.5]{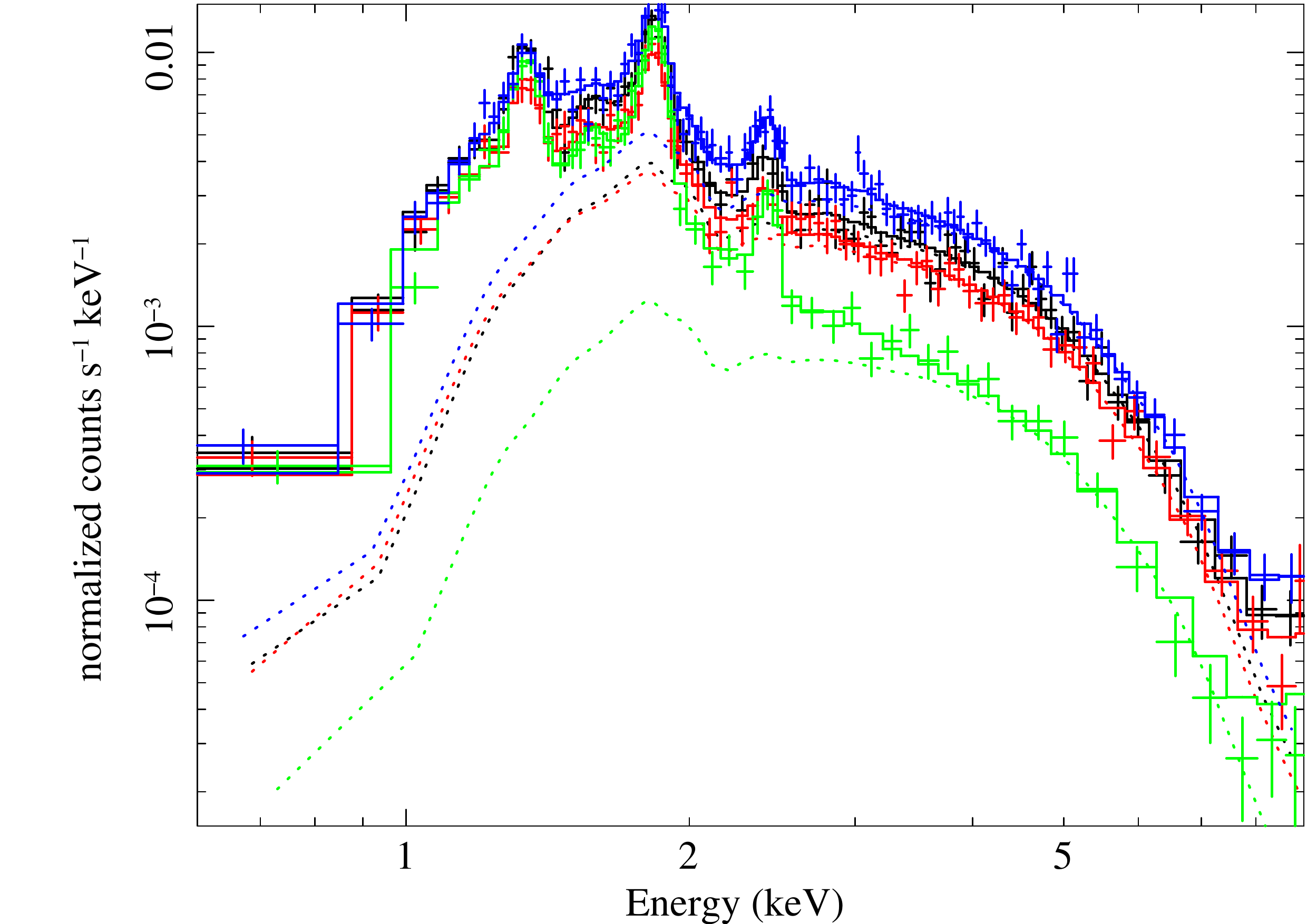}
  \caption{Spectra of the X-ray jets extracted from regions shown in
    Figure~\ref{rxtori} (from NE to SW: blue, red, black, and green),
    with model fits indicated (total -- solid line, jet emission -- dashed
    line). Background has been modeled, not subtracted.
  }
  \label{pwnspectra}
  }
  \end{figure}

\begin{figure}
  \center{
  \includegraphics[angle=0,scale=0.52]{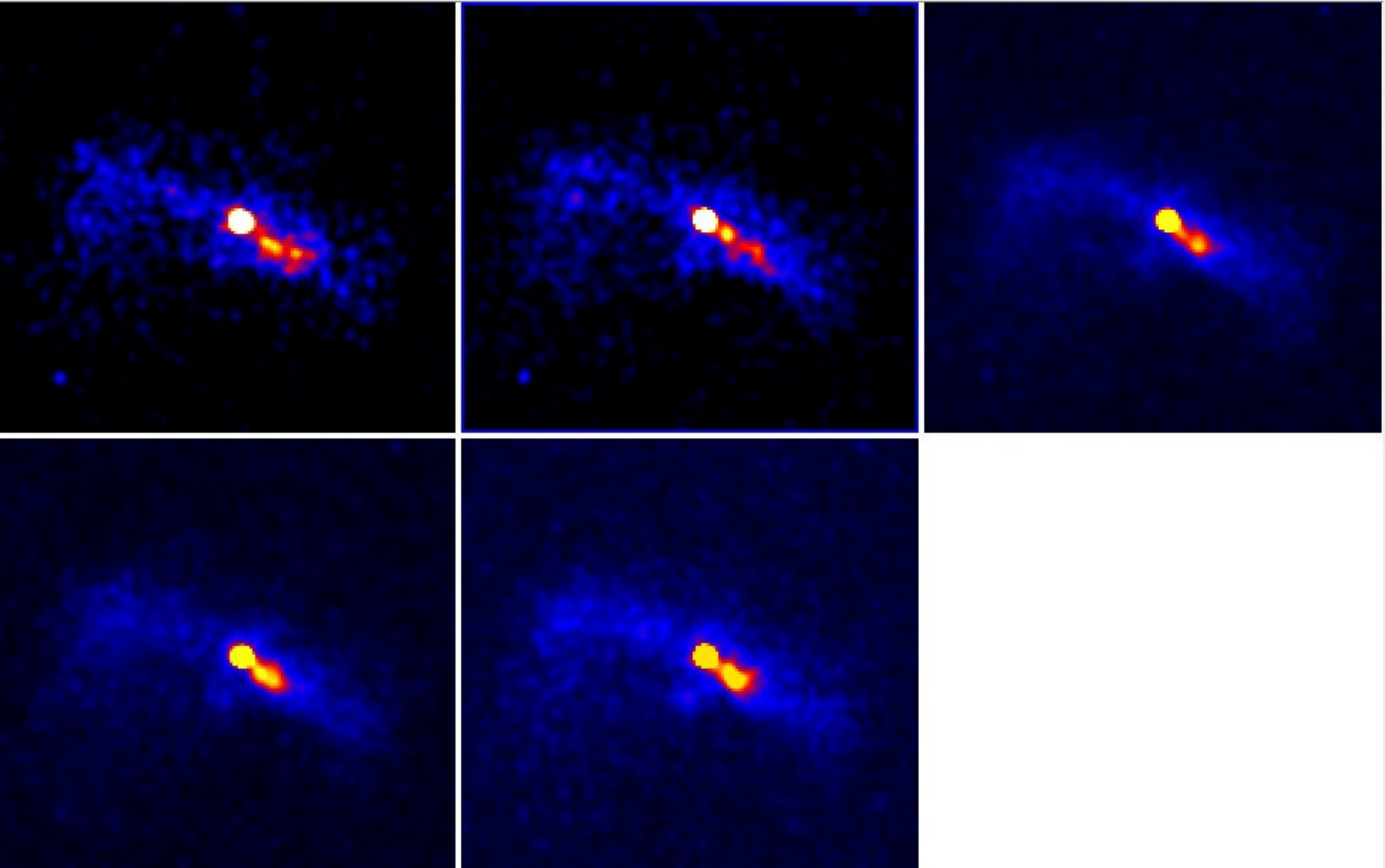}
\caption{X-ray PWN (3.3 -- 8.1 keV) at five epochs.  Top row: 2000,
  2003, 2013 May 5 -- 7; bottom row, 2013 May 25 -- 27 and 2013
  September.  All images are $1.5'$ on a side, and were smoothed
  with a $1.5''$ Gaussian kernel.}
\label{5epochs}
}
\end{figure}

\begin{figure}
  {\hskip-0.5truein\includegraphics[angle=0,scale=0.54]{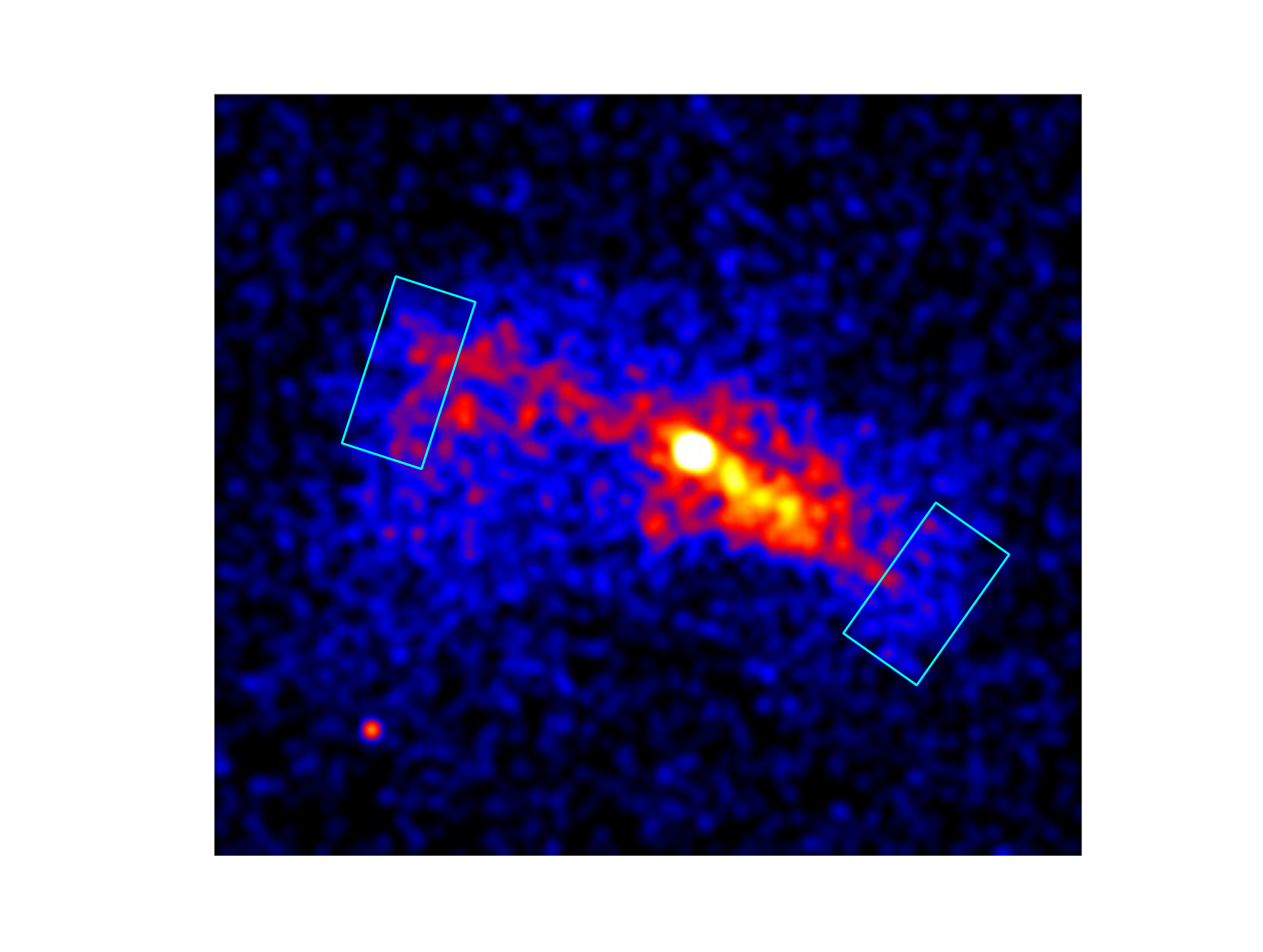}
  \hskip-0.8truein\includegraphics[angle=0,scale=0.54]{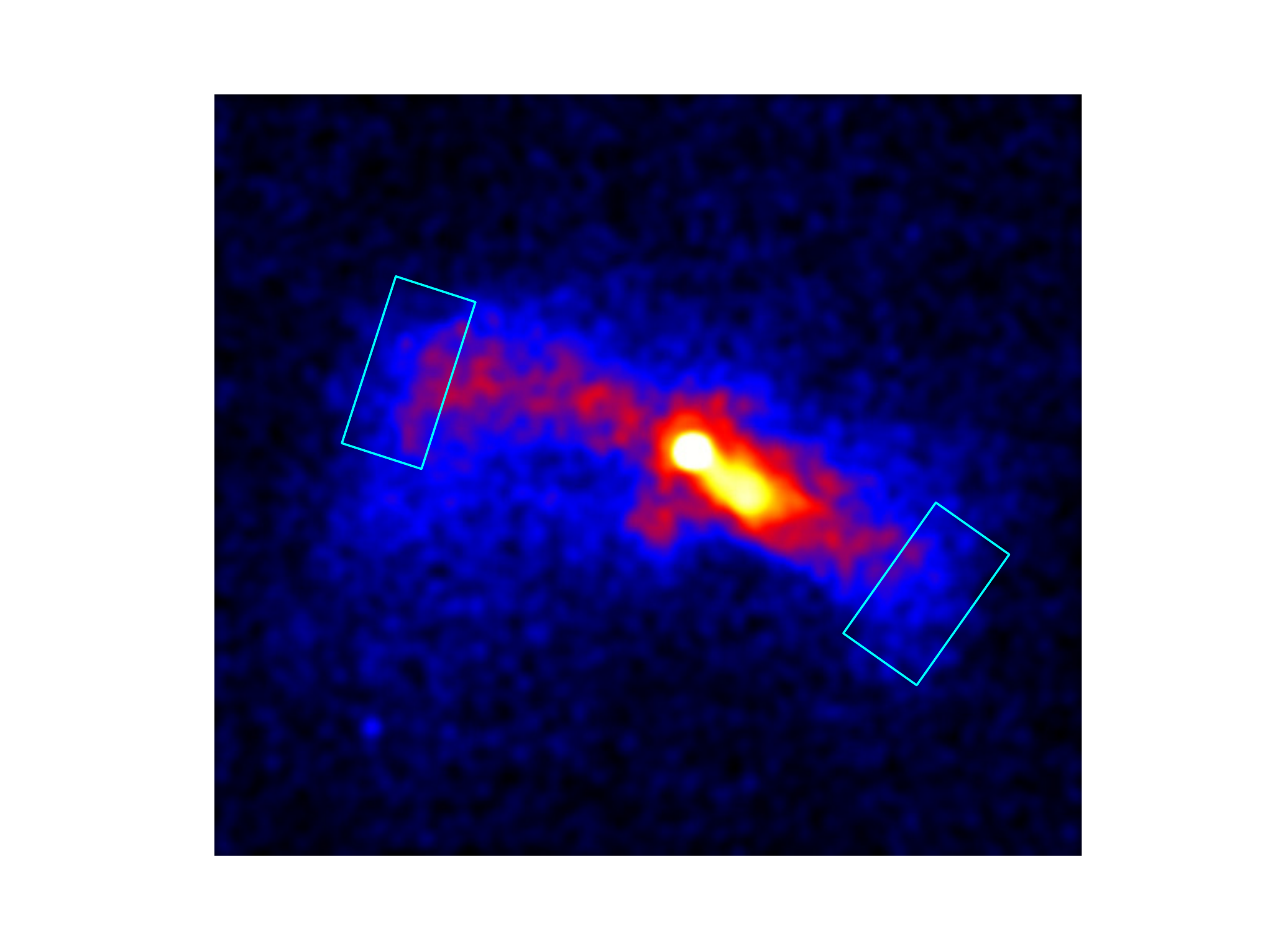}}
  \caption{PWN in 3.3--8 keV X-rays in 2000--2003 and 2013 (left and right
    panels, respectively). Over this period of time, the ends of the jets 
    (enclosed in boxes) receded inward instead of expanding (see text for
    more details). 
  \label{pwnmotion}
  }
  \end{figure}

In Figure~\ref{rxtori} we compare the structures of the X-ray and radio
PWN. While the outer parts are clearly different, both show small scale
structure near the pulsar which are similar in extent and overall
shape, and are reminiscent of the torus+jet structures seen in the
X-ray emission of other PWNe \citep[e.g.,][]{ng04}. What is extraordinary
is that these structures are seen in radio as well as in X-rays.
Figure~\ref{radiotorus} shows a close-up of this radio emission, with
X-ray contours. There
is no evidence of point source emission $> 0.1$~mJy, consistent with
the pulsed upper limits. The putative radio torus is inclined to the
line of sight, with a semi-major axis of  $\sim 12^{\prime \prime}$ and
an apparent axial ratio of about 2 to 1; a circle in
projection would have an axis tilted at $60^\circ$ to the plane of the
sky. The breadth of the torus
appears quite narrow, and is unresolved  in the 3.5~cm image produced
with Robust weighting. To better constrain the thickness of the torus,
we created an image with uniform weighting which is less sensitive to
larger structures but has a synthesized beam of  $1.8^{\prime \prime}
\times 1.3^{\prime \prime}$ where the torus thickness may be barely
resolved. We put an upper limit on the size of the brightest knots in
the torus of $\sim 2^{\prime \prime}$ corresponding to a physical size
of $\sim 0.05$~pc at the nominal 5~kpc distance. The radio image also
shows a clear inner jet-like structure to the Southwest, with the
brightest spot coincident with the brightest time-averaged emission in
the X-rays.

The X-ray jet contains bright knots to the SW (more clearly visible in
Fig.~\ref{5epochs}), while the surface brightness is more uniform to
the NE.  Figure \ref{jetprofile} shows that the mean brightness of the
knots is about three times that of the NE jet. These knots have a hard
spectrum with a photon index $\Gamma$ of $1.36^{+0.10}_{-0.16}$
(errors are 90\%\ confidence intervals). Spectra of selected jet
regions (in magenta in Fig.~\ref{rxtori}) are shown in
Figure~\ref{pwnspectra}. At low energies, these spectra are dominated
by thermal emission produced in the interior of G11.2$-$0.3 (and
discussed in the next section), while at high energies, nonthermal
emission from the jets swamps the much softer thermal interior
emission. We used a plane shock with solar abundances together with a
power law to model these spectra (plus an additional multicomponent
model for the background).  The model fits shown in Figure
\ref{pwnspectra} are the result of a joint fit to all four spectra
with the power-law index $\Gamma$ assumed constant along the jets, but
allowing for variations in absorption. The value of $\Gamma$ is $1.78
\pm 0.07$, so overall the PWN has a significantly softer spectrum than
the bright knots near the pulsar torus. With $\Gamma$ allowed to vary
among regions, we obtain $1.75_{-0.09}^{+0.10}$,
$1.76_{-0.11}^{+0.12}$, $1.79_{-0.11}^{+0.11}$, and
$1.91_{-0.22}^{+0.26}$ (moving from the NE to the SW ends of the jet),
with the C-statistic value decreasing by only 1.6 relative to the fit
with constant $\Gamma$ throughout the jets. Applying the likelihood
ratio test \citep{cash79}, we find that variations in $\Gamma$ are not
statistically significant, so there is no spectral softening along the
X-ray jets as one moves away from the pulsar.  This constrains the
magnetic-field strength and age of the jets (which may be less than
the age of \src).  Absence of softening is also consistent with the
comparable extents of radio and X-ray nebulae.

The brighter structures in the PWN clearly change on a timescale of
weeks to years (Fig.~\ref{5epochs}).  No systematic motion is
apparent, however; the bright knots remain in roughly the same
position while substantial brightness changes occur.  The ends of the
jets do not appear to expand over the 13-year period covered
(Fig.~\ref{pwnmotion}). We measured expansion of the end of the NE jet
using the same method as for expansion of the shell (see
\S~\ref{expansion} for more details), except that here expansion is
measured relative to the pulsar, and we used only high-energy (3.3--8
keV) photons in order to avoid contamination from soft thermal
emission that becomes important at low energies
(Fig.~\ref{pwnspectra}). The end of the NE jet contracted inward
(instead of expanding) by $3.1 (1.4, 4.6)$\%\ (errors are
90\%\ confidence intervals) during the 11.6 yr that separate the mean
epochs of the images shown in Figure~\ref{pwnmotion}. (The quoted
errors do not include systematic effects introduced by smoothing of
long observations from 2013, so this contraction might be
underestimated.)  The contraction rate is $0.27 (0.12,
0.40)$\%\ yr$^{-1}$, or $0.09 (0.04, 0.13)$ arcsec yr$^{-1}$ at the
sharp end where the NE jet terminates. There are not enough high
energy photons at the end of the SW jet to measure its motion
reliably, but a joint fit to both ends of the jets, using regions
shown in Figure~\ref{pwnmotion} and assuming a uniform expansion,
yields contraction of $3.2 (1.8, 4.3)$\%, fully consistent with
contraction of both jets between 2000--2003 and 2013. Because of the
poor photon statistics, it is not clear at this time whether this
apparent contraction is caused by systematic long term motions of the
ends of the jets, or whether it reflects more complex temporal and
morphological variability similar to what is found closer to the
pulsar.  Long-term variations in the pulsar activity might be
responsible for this contraction, or the jets might be
hydrodynamically unstable.  Alternatively, since, as we argue
below, the reverse shock has already arrived at the center
and reheated all the ejecta, this interaction may have
produced more complex PWN dynamics.

\begin{figure}
  \center{
  \includegraphics[angle=270,scale=0.5]{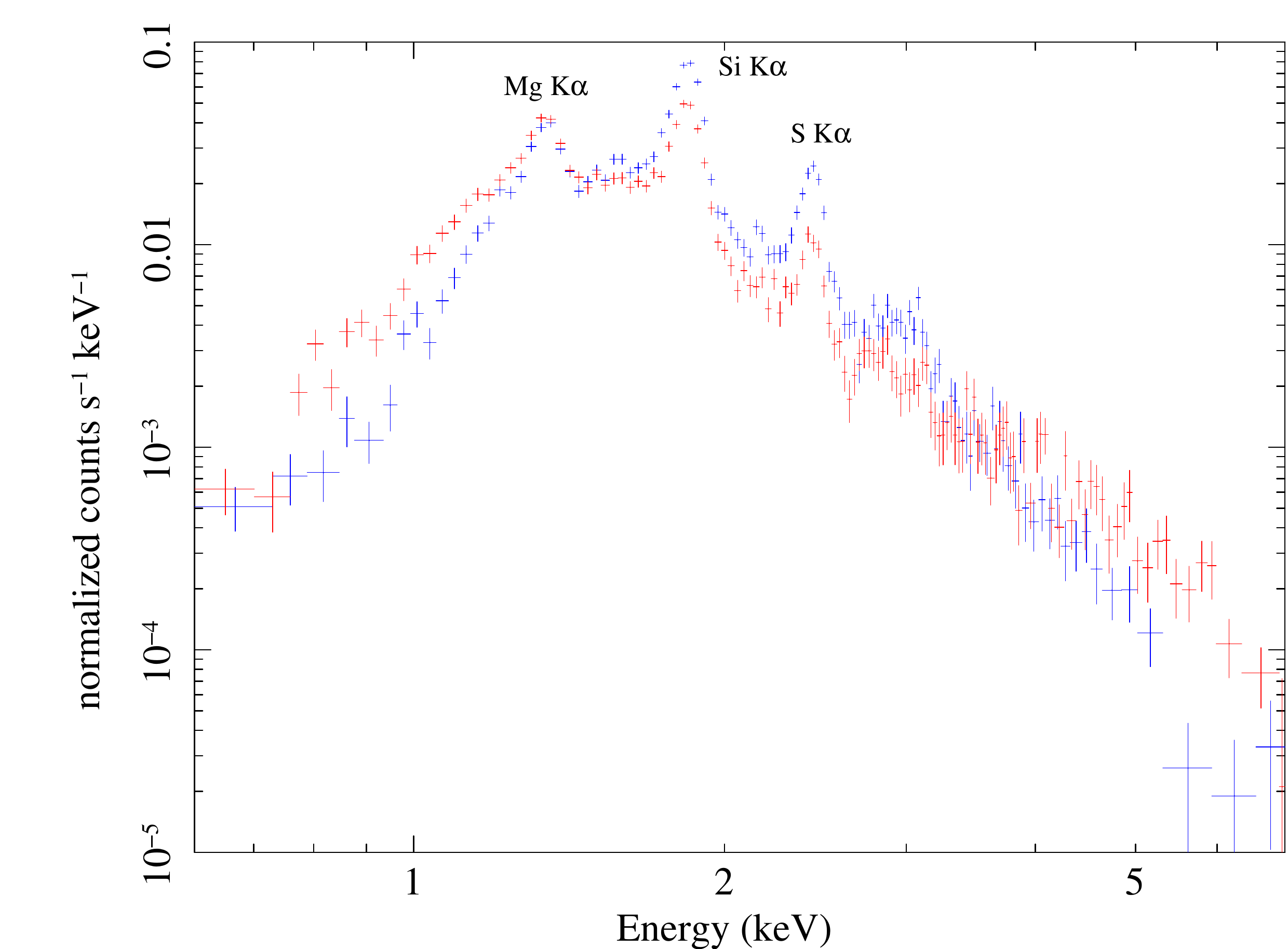}
\caption{Spectra of the NW shell region (blue) and the central bar
  (red; excluding the X-ray jet, see Figure \ref{shell13}).
  Note that the central emission is
  dominantly thermal, showing the same clear lines of He-like Mg, Si,
  and S as the outer shell, but with a softer spectrum overall but
also a hard excess.  We interpret the hard excess as emission from
the fainter regions of the PWN overlapping the bar.}
\label{2specs}
}
\end{figure}

\begin{figure}
  \center{
  \includegraphics[angle=0,scale=0.4]{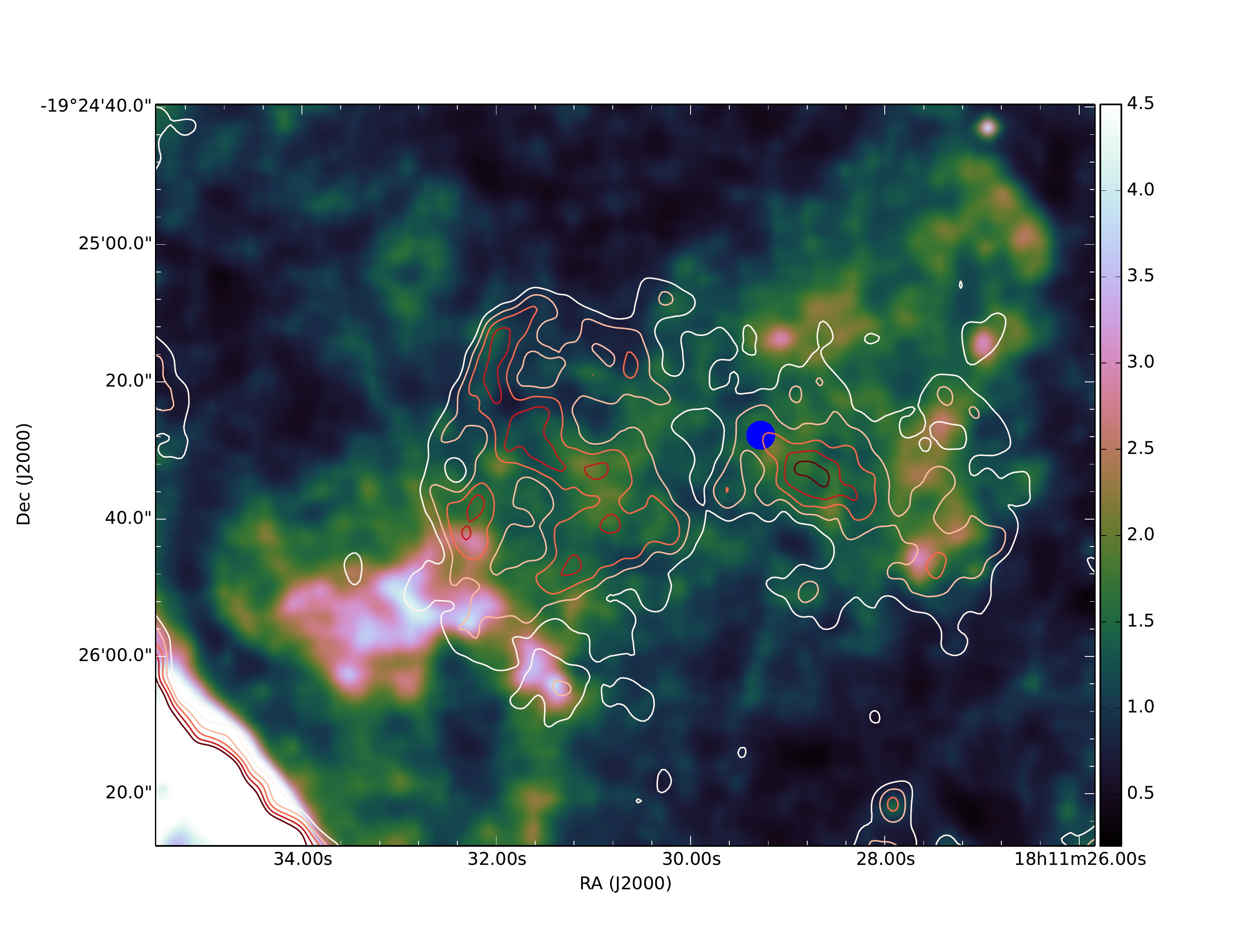}
\caption{X-ray interior emission in the Mg K$\alpha$ line, overlaid 
with 3.5 cm radio contours. Scale is in counts per $0.6'' \times
0.6''$ image pixel in the 1.29--1.40 keV energy range. Radio contours
range from 1 mJy beam$^{-1}$ to 1.7 mJy beam$^{-1}$.  Note how the
radio PWN fills the gap in the bar of soft X-ray emission extending
across the remnant center.}
\label{r-cavity}
}
\end{figure}

\begin{figure}
\epsscale{1.1}
\vspace{0.1truein}
\plotone{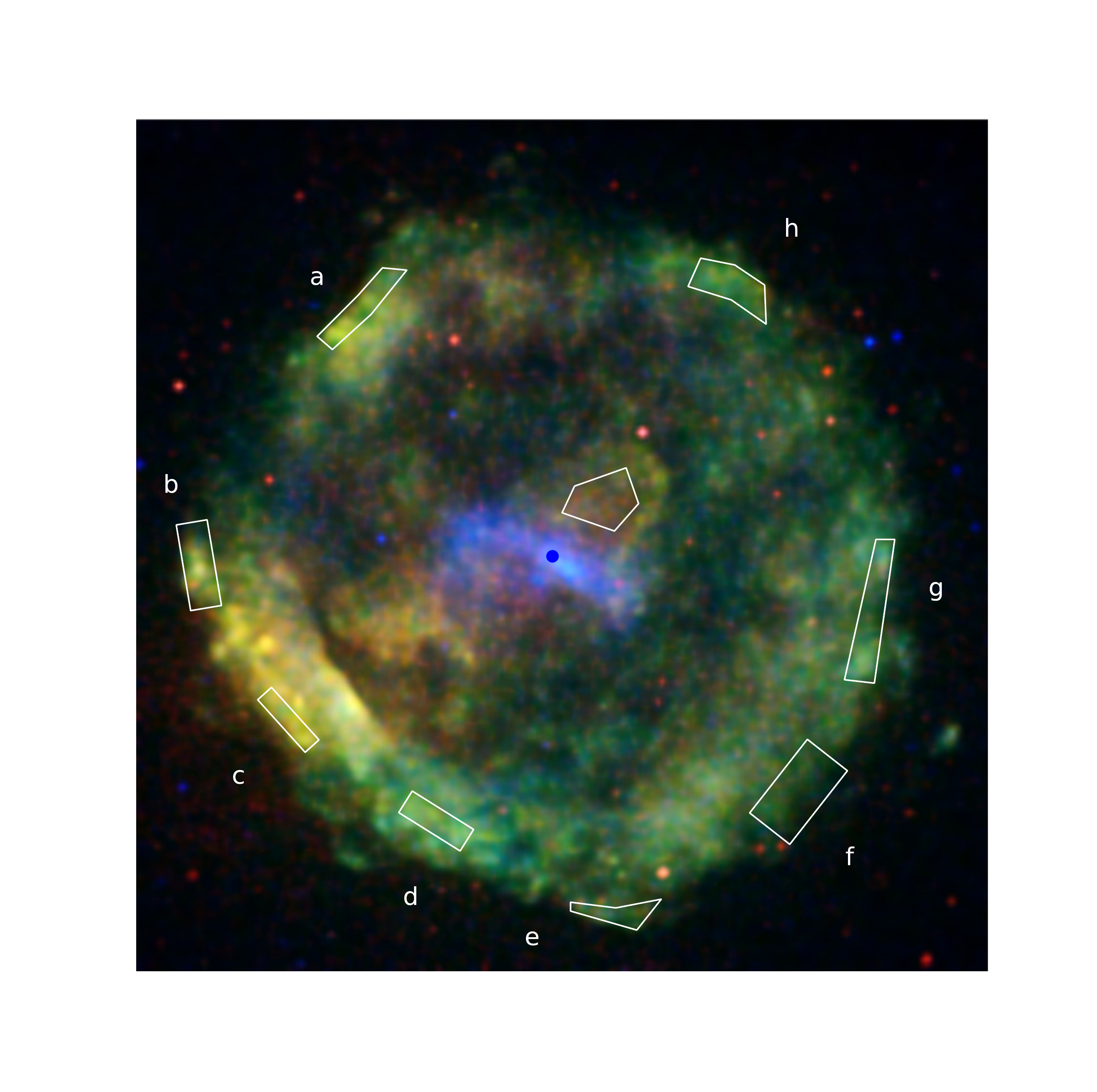}
\caption{
2013 {\sl Chandra} image of G11.2$-$0.3 \citep[image was smoothed 
with the multispectral method of][]{salmon14}, with shell regions chosen for
the spectral analysis overlaid. Red:
0.6 -- 1.2 keV; green, 1.2  -- 3.3 keV; blue, 3.3 -- 8 keV.
The central blue circle marks the location of the (masked) pulsar.
Soft X-ray emitting region near the pulsar is also shown.
Image size $308'' \times 308''$. 
\label{shell13}}
\end{figure}

\section{Thermal and Nonthermal Interior Emission}

\src\ contains an anomalous bar of interior X-ray emission, running
roughly from SE to NW.  Its spectrum, from a region close to the PWN
but excluding the X-ray jets (and shown in the center of
Fig.~\ref{shell13}), is compared with the NW region of the shell in
Figure \ref{2specs}.  It is clearly thermal, with a soft excess
compared to the shell, but also with excess emission above about 3
keV, which we attribute to contamination from the outer, fainter parts
of the PWN.  The PWN appears to occupy a cavity in that bar of thermal
emission.  Figure \ref{r-cavity} compares the soft X-ray image in the
Mg K$\alpha$ line with the 3.5-cm radio image.  A brighter rim can be
seen at the SE edge of the PWN.  Comparison of the X-ray PWN with the
thermal X-ray image (Fig.~\ref{shell13}) confirms this conclusion.

\begin{figure}
  \center{
\includegraphics[angle=0,scale=0.42]{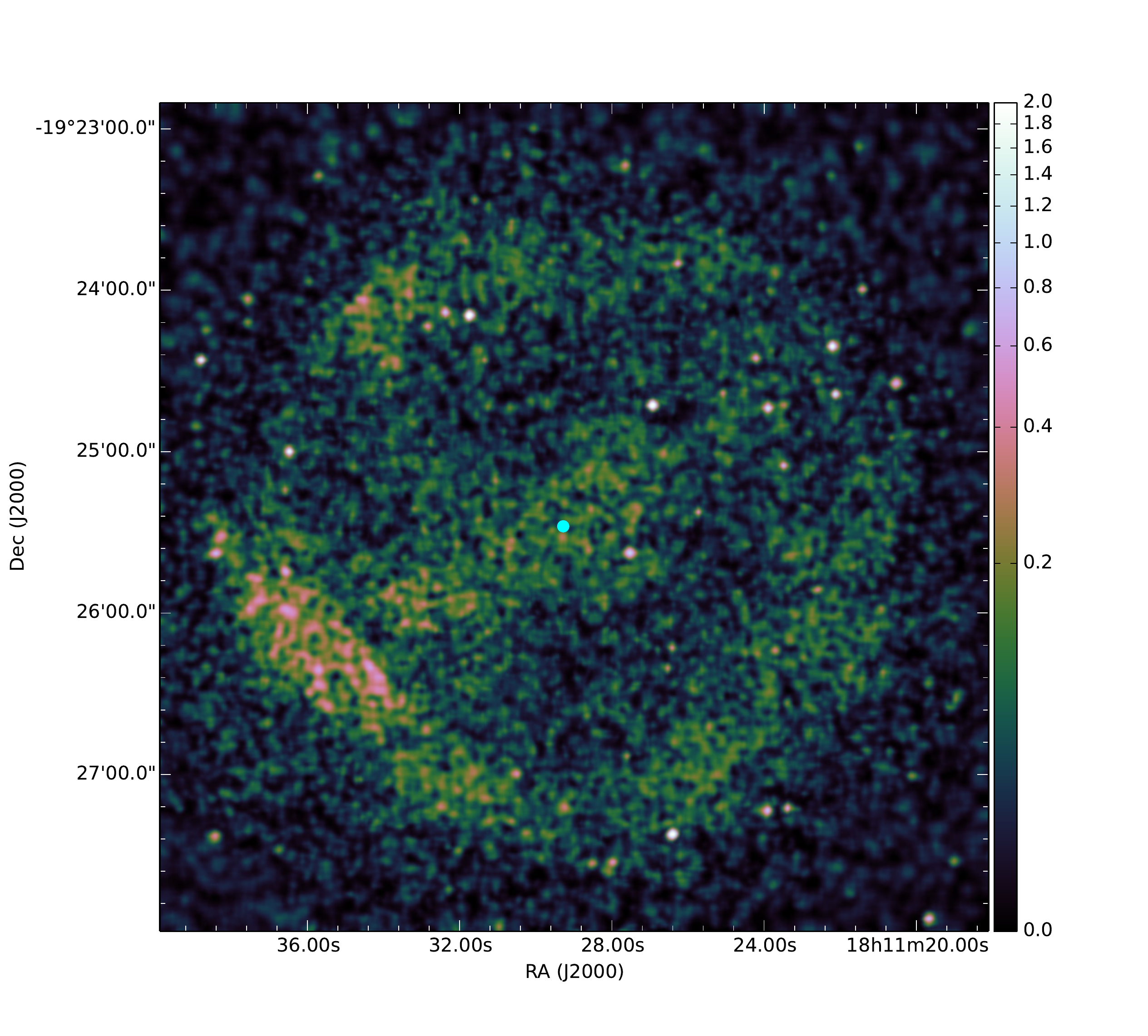} 
\includegraphics[angle=0,scale=0.42]{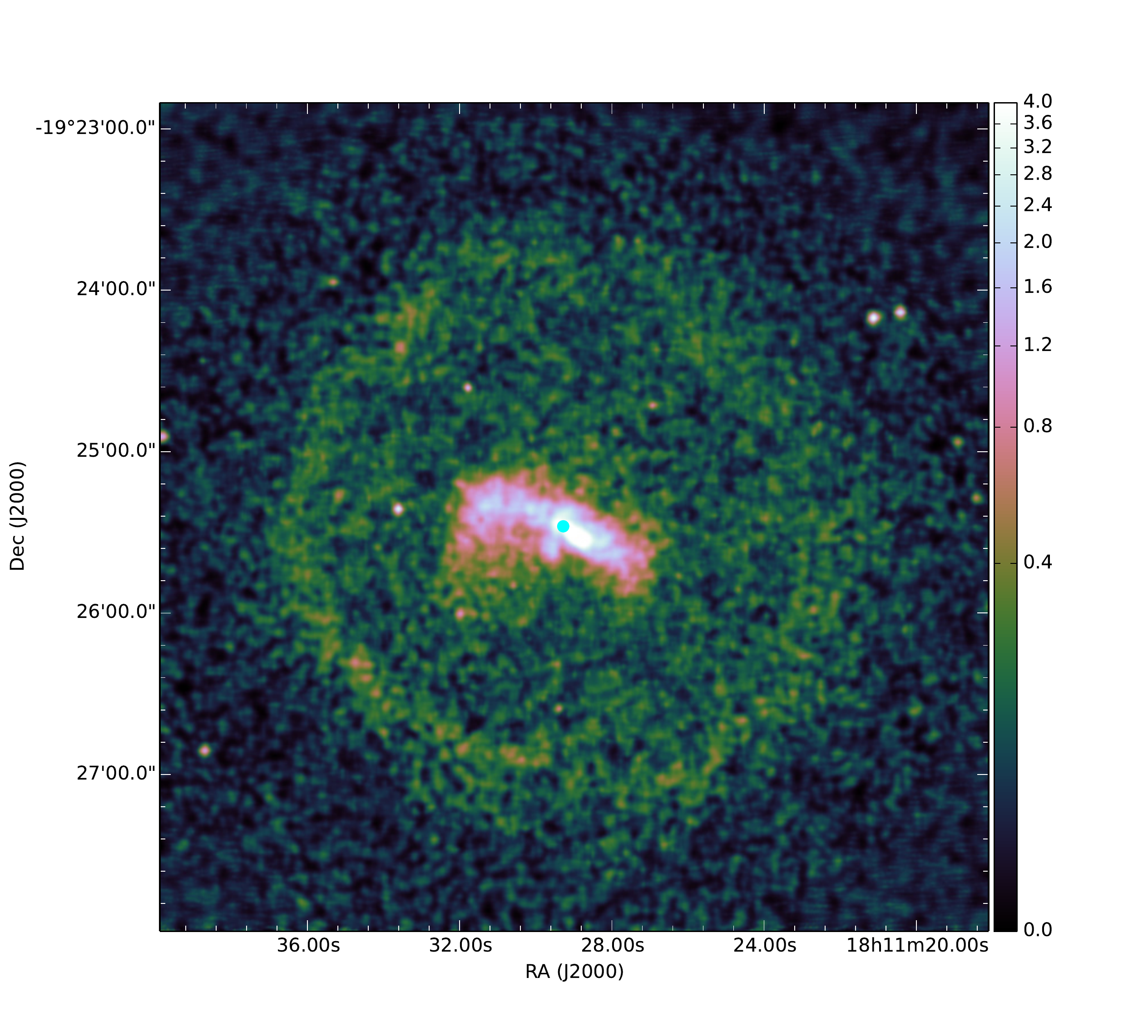}
\caption{Smoothed images of G11.2$-$0.3 in very soft
  (0.59 -- 0.82 keV; top) and very hard (5 -- 8 keV; bottom) X-rays.
  Scale is in counts per
  $0.6'' \times 0.6''$ image pixel. The pulsar has been masked out (cyan
  regions). The brightest parts of the PWN are saturated. 
  Note striking difference in the shell radius between soft and hard X-rays.
  }
\label{figsofthard}
         }
\end{figure}

The clear interaction of the PWN with the thermal emission indicates
that that material is in the remnant interior, and not just projected
from the front or back face of the shell.  That is, material all the
way in to the PWN has been reheated by the reverse shock.  This is an
important clue to the evolutionary state of \src, indicating that
that state is quite advanced, and suggesting that nearly all the SN
ejecta have been shocked by now.

We find no evidence for enhanced abundances in the interior of
G11.2$-$0.3.  While there is a soft excess in the central bar spectrum
(Fig.~\ref{2specs}), the abundances of Mg, Si, and S whose lines
dominate X-ray spectra are not appreciably different than elsewhere in
the remnant.  It is likely that the X-ray spectrum of the central bar
is dominated by the swept-up ambient gas and not the SN ejecta. In
view of the advanced evolutionary stage of the remnant, the shocked
ejecta must be present there as well, but apparently they do not emit
as efficiently as the swept-up ambient medium (at least at X-ray
energies high enough for photons to effectively penetrate the
intervening ISM).

Contrasting morphologies of G11.2$-$0.3 in very soft (0.59 -- 0.82
keV) and very hard (5 -- 8 keV) X-rays are shown in Figure
\ref{figsofthard}. The very soft, predominantly thermal X-ray emission
at the center lacks limb-brightened rims seen in the Mg K$\alpha$ line
(Fig.~\ref{r-cavity}), while at very high energies nearly all
X-rays there come from the PWN and the pulsar. There is a shell
visible in hard X-rays but its diameter is substantially smaller than
the overall X-ray size of G11.2$-$0.3. Within this shell, hard diffuse
emission fills the remnant's interior. This hard X-ray shell roughly
coincides with the inner edge of the remnant's shell, but does not
match it well on small spatial scales. Apparently, this interior
emission is produced within a spatially-distinct region within
G11.2$-$0.3. In order to produce such hard ($> 5$ keV) photons, either
hot thermal plasma with a temperature of several keV or nonthermal
X-rays are required. There is no trace of the Fe K$\alpha$ line in
X-ray spectra, so it is possible that this emission is of nonthermal
origin. But irrespective of its (thermal or nonthermal) origin, the
presence of a very hard X-ray shell in the interior of a dynamically
evolved remnant such as G11.2$-$0.3 is quite unexpected.

\section{G11.2$-$0.3 and a Possible SN in CE 386}

\citet{lee13} found very high extinction ($A_V=16^m$--$20^m$) in their
study of 3 IR-bright rims in the SE and S. At 5 kpc distance,
the SN would have been fainter than $29^m+M_V$,
clearly not compatible with historical reports of a possible SN in CE
386 for even the most luminous SNe.  A typical CC SN ($M_V=-18^m$)
could not have been seen by naked eye even if $A_V$ were as low as
$12^m$ reported by \citet{lee13} in a few spatially localized regions
in the SE. We demonstrate here that absorption is high not only in the
SE and S, ruling out association of G11.2$-$0.3 with a possible SN
386.

\begin{figure}
  \center{
\epsscale{0.9}
\vspace{0.1truein}
\plotone{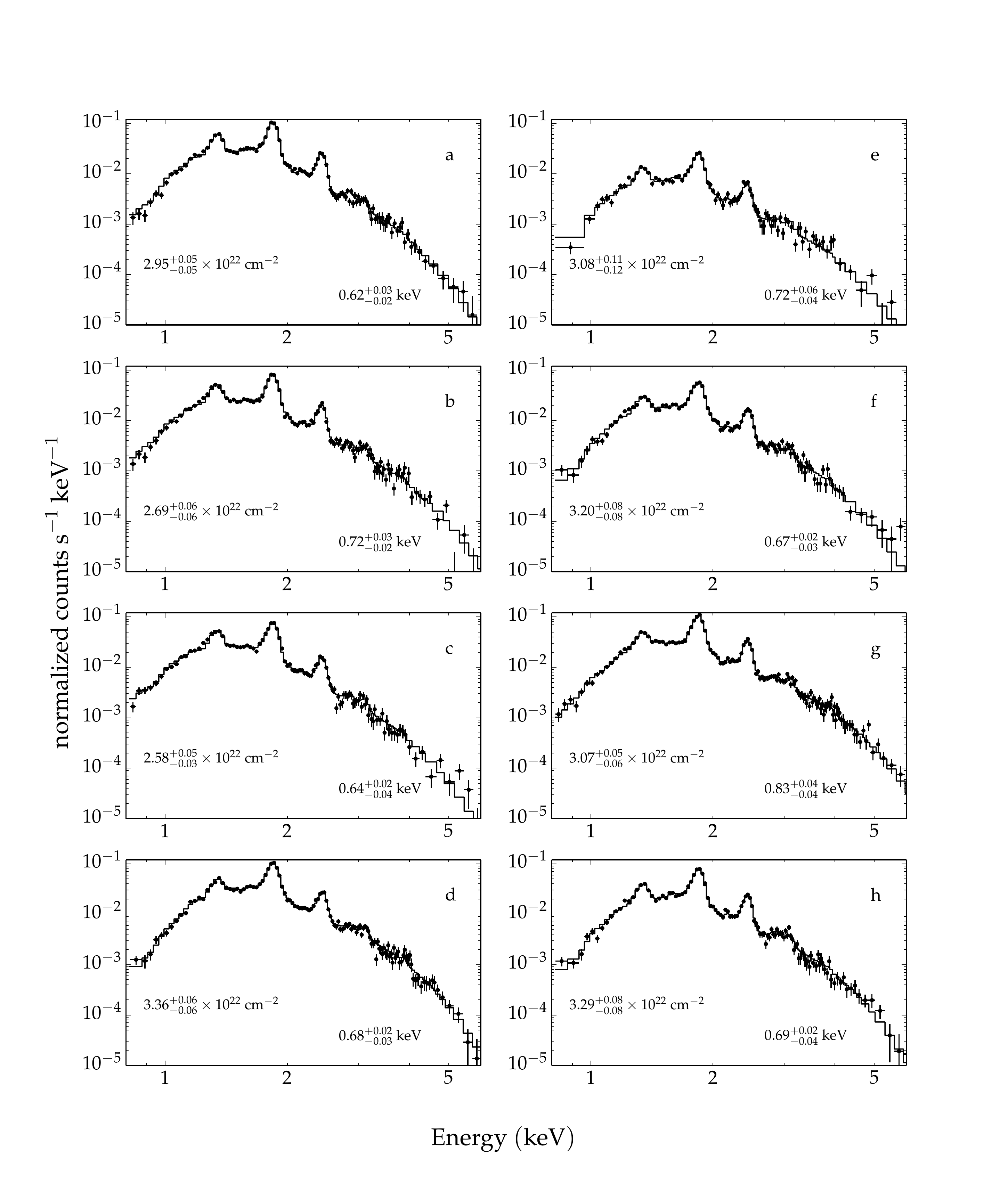}
\caption{
  Spectra of the shell in regions shown in Figure \ref{shell13}, overlaid with
  model spectra. For each spectrum, the fitted values of hydrogen column
  density $N_H$ and electron temperature $T_e$ are shown (errors are
  90\%\ confidence intervals).
  \label{spectra13}}
}
\end{figure}

We analyzed X-ray spectra of 8 outer shell sections (see
Fig.~\ref{shell13}) by fitting them with a plane shock model ({\tt 
vpshock} model in XSPEC) with atomic data from AtomDB
\citep{smith01,foster12}. We assumed solar abundances as given by
\citet{grsa98} except for Mg, Si, and S whose abundances have been
allowed to vary in our fits.  As shown in Figure \ref{spectra13}, the
fitted values of hydrogen column density $N_H$ vary from $2.6 \times
10^{22}$ cm$^{-2}$ to $3.4 \times 10^{22}$ cm$^{-2}$, with the least
amount of absorption found in the E and SE (regions {\sl b} and {\sl 
c}, consistent with the soft X-ray excess seen there in Figure
\ref{shell13}). (Near the pulsar, we find slightly lower
($N_H = 2.4$--$2.7 \times 10^{22}$ cm$^{-2}$) absorption from fits to the spectra
in Figure \ref{pwnspectra}, but we consider them less reliable in view of
the mixed thermal and nonthermal nature of these interior spectra, and with
the thermal interior emission not as well understood as the outer shell
emission.)
In order to convert from $N_H$ to $A_V$, we assume the
same dust-to-gas mass ratio and the same dust properties as in the
dust model of \citet{wg01} with $R_V=3.1$, so $A_V=15\fm9(N_H/3 \times
10^{22}{\rm cm}^{-2})$. This gives $A_V$ in the range from $14^m$ to
$18^m$ across the entire remnant, confirming the high absorption
derived from IR observations.

Our optical extinction estimates depend on a number of uncertain
assumptions, but this is unlikely to affect our conclusions about
the non-association of G11.2$-$0.3 with a possible SN 386. We assumed
solar abundances of \citet{grsa98} for both absorbing and X-ray
emitting gas, the same solar abundance set used by \citet{wg01} in
their dust models constrained by observations of dust in the solar
vicinity. Since in the inner Galaxy metal abundances are higher than
in the solar neighborhood, this leads to an overestimate of $N_H$ (at
the high absorption of interest here, heavy elements alone, and not H and
He, are responsible for X-ray absorption). But for the same degree of
heavy element depletion onto dust expected in the diffuse ISM across
the entire Galaxy, $A_V$ remains the same.

Our neglect of X-ray scattering by dust leads to overestimation of
both $N_H$ and $A_V$ but this effect is difficult to quantify for
spatially-extended sources such as G11.2$-$0.3. \citet{reynolds09}
found a 25\%\ reduction in $N_H$ for the much more compact and much
more heavily absorbed ($N_H=5.1 \times 10^{22}$ cm$^{-2}$) remnant
G1.9+0.3 by modeling its dust-scattered X-ray halo. But even such a
large and unrealistic reduction in $A_V$ would not make a typical CC
SN visible without a telescope at 5 kpc distance. Most likely, this
reduction is quite modest for G11.2$-$0.3, as our estimates of $A_V$
in regions {\sl c} and {\sl d}, $14^m$ and $18^m$, do not exceed the
$16^m$--$18^m$ found by \citet{lee13} in this region of the
remnant. (In principle, IR-based estimates of optical extinction
should be more reliable than our X-ray-based estimates, but currently
there are still uncertainties in atomic data for the [\ion{Fe}{2}]
transitions used by \citet{lee13} to estimate IR extinction
\citep{giannini15,koolee15}.)

Recent SNe within the inner Galaxy, such as those that produced
G11.2$-$0.3 and the youngest known Galactic SNR G1.9+0.3, were too
heavily absorbed to be detected at optical wavelengths. If the guest
star of 386 were indeed a supernova, it must have been not so distant
and heavily absorbed. Lack of a suitable candidate remnant associated
with this guest star casts serious doubts about its identification as
a supernova event.

\section{Shell Expansion} \label{expansion}

\begin{figure}
\epsscale{0.82}
\vspace{0.1truein}
\plotone{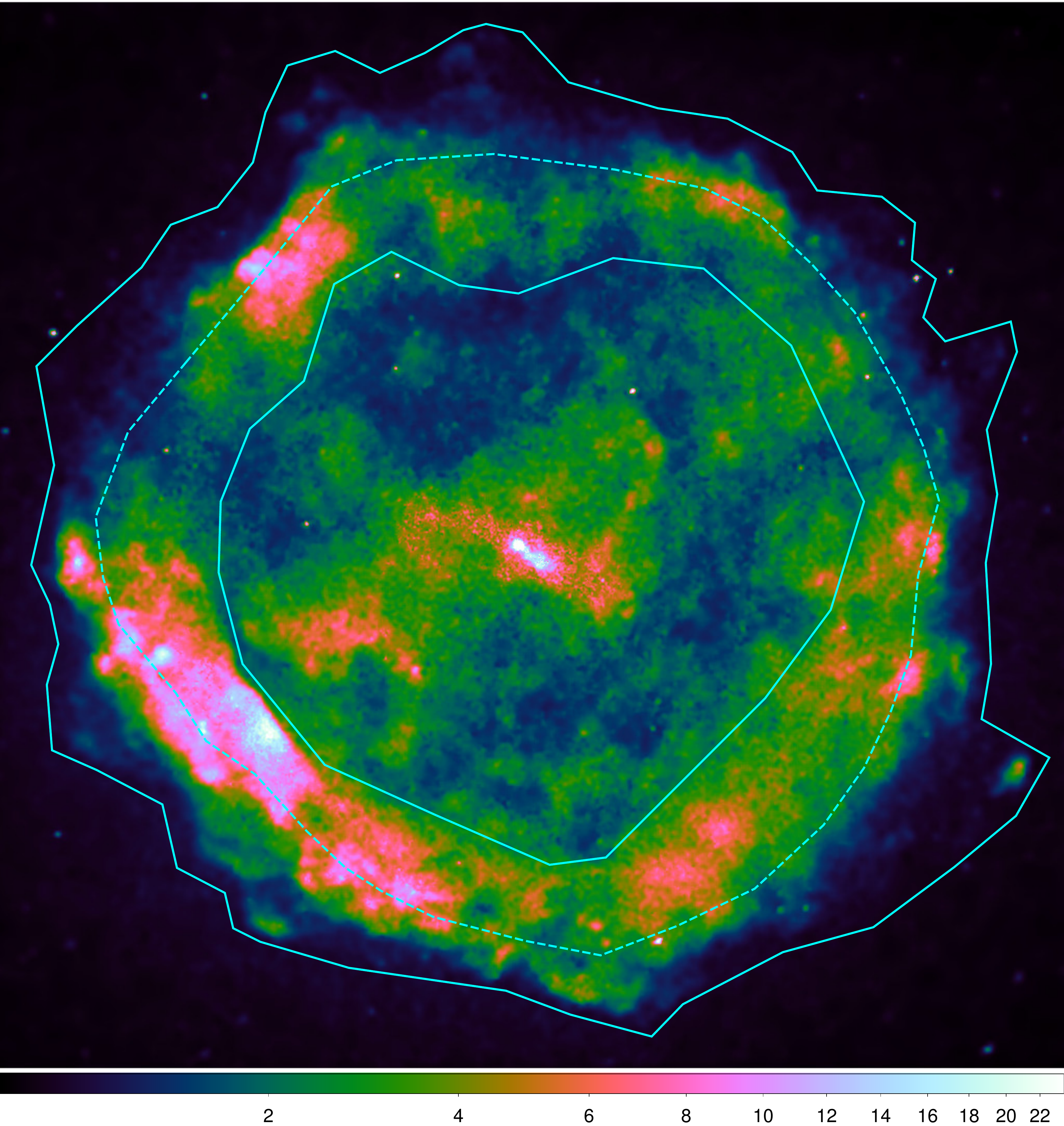}
\caption{
2013 May {\sl Chandra} image of G11.2$-$0.3. Scale is in 
counts per $0.3'' \times 0.3''$ image pixel in the 0.6--8 keV energy 
range \citep[image was smoothed 
with the multiscale partitioning method of][]{krishnamurthy10}. Shell regions 
chosen for expansion measurements are shown.
Intensities shown with the cubehelix color 
scheme of \citet{green11}. 
Image size $308'' \times 308''$. 
\label{ximmay}}
\end{figure}

{\it Chandra} observed \object{G11.2$-$0.3} in 2000 (Epoch I
observations), 2003 (Epoch II), and 2013 (Epoch III), over a timespan
of nearly 13 yr. This time baseline is long enough to measure
expansion of the remnant reliably. We first measured the overall
expansion using a variation of the method described in
\citet{carlton11} \citep[see also][]{vink08}. From the merged 2013 May {\sl 
Chandra} data, we extracted a spectral cube, $1024^2 \times 128$ in
size, that included counts from the entire remnant in the 0.6--8 keV
energy range (spectral channels 41--552). Event positions were binned
to about $3/5$ of the ACIS pixel size, so one image pixel is $0.301''
\times 0.301''$. With $1024^2$ spatial pixels, the $5.1' \times 5.1'$
area encompasses the entire remnant, including the outlying outer
knots.  Spectral channels were binned by a factor of 4. The 2013 May
data cube was then smoothed with the spectro-spatial method of
\citet{krishnamurthy10}, varying a penalty parameter that controls the
amount of smoothing. Smoothed datacubes were summed over spectral
channels to yield smoothed 0.6--8 keV images.  We used these images as
a model for the surface brightness of \object{G11.2$-$0.3} at Epoch
III (see Figure \ref{ximmay} for a model with the penalty parameter of
0.015).  These model images (after background subtraction) were then
fit to the unsmoothed 0.6--8 keV images from earlier epochs (i.e.,
shrunk to fit) using the maximum likelihood method
\citep[C-statistic;][]{cash79}.  We excluded the remnant's interior in
these fits as shown in Figure \ref{ximmay}. We also excluded poorly
exposed outer sections of the shell at Epoch II. Several point sources
were masked out. Spatial variations in effective exposure times were
taken into account by correcting model images with the help of
monochromatic ($E = 1.7$ keV) exposure maps.  For each observation
from Epochs I and II, there are four free parameters in this model: a
physical scaling factor, a surface-brightness scaling factor, and
image shifts in right ascension $\alpha$ and declination
$\delta$. These image shifts allow correction of any misalignment of
individual pointings due to the {\sl Chandra} external astrometric
errors \citep[mean error of $0.''16$;][]{rots09}. They might also be
nonzero for perfectly aligned observations in the presence of 
asymmetric expansion. In this case, expansion measurements reported
here provide us with the mean expansion rate of the remnant.

Since our model images only approximate the true surface brightness
distribution, model uncertainties contribute to errors in measured
expansion rates. We took these systematic effects into account by
using the 0.6--8 keV image from 2013 September as the control
image. In the 0.32 yr time baseline from 2013 May to September, we
expect a negligible (only $\sim 0.01$\%) expansion of the remnant. But
fits to the 2013 September image, allowing only for changes in the
physical scaling and surface brightness factors, revealed systematic
trends with the amount of smoothing. Increased smoothing broadens the
outer boundary of the remnant, so the model must be shrunk to match
unsmoothed data. As the magnitude of this systematic effect is
expected to be dependent on the model alone, the physical scaling
factor derived from fitting the 2013 September image was used to
correct the measured Epoch I and II scaling factors. We verified that
the corrected scaling factors showed no significant variations with
the amount of smoothing (except for either grossly inadequate or
excessive smoothing).

\begin{deluxetable}{lccccccc}
\tablecolumns{8}
\tablewidth{0pc}
\tablecaption{Expansion of G11.2$-$0.3}

\tablehead{
\colhead{Baseline}  & Observation ID & $\Delta t$\tablenotemark{a} & $\Delta \alpha \cos \delta$\tablenotemark{b} & $\Delta \delta$\tablenotemark{c} & $S$\tablenotemark{d} & Expansion\tablenotemark{e}  & Expansion Rate\tablenotemark{e} \\
\colhead{} & & (yr) &\multicolumn{2}{c}{(arcsec)} & & (\%) & (\%\ yr$^{-1}$)}
\startdata
%\\

2013 May -- 2013 Sep & \nodata & 0.32 & \nodata & \nodata & $1.006 \pm 0.002$ & $0.007 \pm 0.014$& \nodata \\

2003 -- 2013 May & 3909 & 10.01 & 0.043 & 0.054 & $1.041 \pm 0.005$ & \nodata & \nodata \\
& 3910 & 9.88 & 0.043 & 0.053 & $1.049 \pm 0.005$ & \nodata & \nodata \\
& 3911 & 9.79 & 0.065 & 0.082 & $1.033 \pm 0.007$ & \nodata & \nodata \\
& 3912 & 9.68 & 0.042 & 0.054 & $1.038 \pm 0.005$ & \nodata & \nodata \\
& \nodata & 9.85 & \nodata & \nodata & \nodata & ($0.268 \pm 0.019$) & ($0.0271 \pm 0.0019$) \\
& \nodata & 9.85 & \nodata & \nodata & \nodata & $0.266 \pm 0.023$ & $0.0270 \pm 0.0024$ \\

2000 -- 2013 May & 780 & 12.77 & 0.038 & 0.046 & $1.046 \pm 0.004$ & \nodata & \nodata \\
& 781 & 12.58 & 0.050 & 0.056 & $1.052 \pm 0.006$ & \nodata & \nodata \\
& 2322 & 12.58 & 0.070 & 0.079 & $1.044 \pm 0.009$ & \nodata & \nodata \\
& \nodata & 12.69 & \nodata & \nodata & \nodata & ($0.361 \pm 0.025$) & ($0.0285 \pm 0.0020$) \\
& \nodata & 12.69 & \nodata & \nodata & \nodata & $0.359 \pm 0.029$ & $0.0283 \pm 0.0023$ \\

(2000+2003) -- 2013 May & \nodata & \nodata & \nodata & \nodata & \nodata & \nodata & $0.0277 \pm 0.0018$
\enddata
\tablecomments{\ All errors are $1\sigma$.}
\tablenotetext{a}{Baseline length.}
\tablenotetext{b}{Alignment error in right ascension.}
\tablenotetext{c}{Alignment error in declination.}
\tablenotetext{d}{Model surface brightness scaling.}
\tablenotetext{e}{Values in brackets are before correction for systematic effects.}
\label{expansiontable}
\end{deluxetable}

The measured expansion rates are listed in Table \ref{expansiontable}
for a model with the penalty parameter of 0.015, together with the
alignment errors, surface scaling factors, and physical scaling
factors. The mean error on displacements among individual observations
is $0.07''$. The surface brightness scaling factors are larger than
unity because we used monochromatic exposure maps that do not take
into account the continuing decrease in time of the {\it Chandra}
low-energy sensitivity.  The chosen penalty parameter leads to a
reasonably smooth model image (Figure \ref{ximmay}) without producing
any appreciable bias in measured scaling factors. The Epoch I and II
observations were fit independently, then corrected for the
bias. Errors were added in quadratures. A larger (by $\sim 30$\%)
expansion is found for the Epoch I observations, in agreement with
expectations based on the lengths of the time baselines. The corrected
expansion rates for Epochs I and II are within measurement
errors. Their weighted average is $0.0277 \pm 0.0018$\%\ yr$^{-1}$.

The inner boundary of the shell is quite sharp in several locations,
offering us an opportunity to search for possible variations in
expansion across the shell. We divided the shell into inner and
outer sections (Figure \ref{ximmay}), and measured their expansion
rates. The model is based on all 2013 data, and we used the same
(0.015) penalty parameter in order to avoid biasing expansion
measurements. The Epoch I and II observations were fit simultaneously,
assuming constant expansion rates. We refit the whole shell again to
verify that this measurement method is consistent with the more
rigorous procedure described above. There is excellent agreement
between the two methods, as the overall shell expansion is $0.0267 \pm
0.0013$\%\ yr$^{-1}$ (Table \ref{expansiontablea}). (Errors listed in
Table \ref{expansiontablea} do not include systematic effects.)

\begin{deluxetable}{lccccc}
\tablecolumns{6}
\tablewidth{0pc}
\tablecaption{Inner and Outer Shell Expansion}

\tablehead{
\colhead{Region}  & Observation ID & $\Delta \alpha \cos \delta$ & $\Delta \delta$ & $S$ & Expansion Rate\tablenotemark{a} \\
\colhead{} & &\multicolumn{2}{c}{(arcsec)} & & (\%\ yr$^{-1}$)}
\startdata

Whole Shell & 3909 & 0.040 & 0.042 & $1.037 \pm 0.005$ & \nodata \\
& 3910 & 0.038 & 0.047 & $1.045 \pm 0.005$ & \nodata \\
& 3911 & 0.063 & 0.074 & $1.030 \pm 0.007$ & \nodata \\
& 3912 & 0.044 & 0.056 & $1.034 \pm 0.005$ & \nodata \\
& 780 & 0.036 & 0.045 & $1.042 \pm 0.004$ & \nodata \\
& 781 & 0.050 & 0.054 & $1.047 \pm 0.006$ & \nodata \\
& 2322 & 0.063 & 0.091 & $1.040 \pm 0.009$ & \nodata \\
& \nodata & \nodata & \nodata & \nodata & $0.0267 \pm 0.0013$ \\
Outer Shell & 3909 & 0.054 & 0.056 & $1.037 \pm 0.009$ & \nodata \\
& 3910 & 0.061 & 0.066 & $1.040 \pm 0.010$ & \nodata \\
& 3911 & 0.075 & 0.082 & $1.042 \pm 0.013$ & \nodata \\
& 3912 & 0.067 & 0.061 & $1.035 \pm 0.009$ & \nodata \\
& 780 & 0.051 & 0.056 & $1.042 \pm 0.008$ & \nodata \\
& 781 & 0.065 & 0.066 & $1.027 \pm 0.011$ & \nodata \\
& 2322 & 0.089 & 0.090 & $1.033 \pm 0.016$ & \nodata \\
& \nodata & \nodata & \nodata & \nodata & $0.0279 \pm 0.0018$ \\
Inner Shell & 3909 & 0.060 & 0.063 & $1.038 \pm 0.006$ & \nodata \\
& 3910 & 0.048 & 0.060 & $1.050 \pm 0.006$ & \nodata \\
& 3911 & 0.126 & 0.076 & $1.028 \pm 0.009$ & \nodata \\
& 3912 & 0.057 & 0.085 & $1.038 \pm 0.006$ & \nodata \\
& 780 & 0.048 & 0.065 & $1.045 \pm 0.005$ & \nodata \\
& 781 & 0.058 & 0.067 & $1.061 \pm 0.007$ & \nodata \\
& 2322 & 0.088 & 0.099 & $1.045 \pm 0.011$ & \nodata \\
& \nodata & \nodata & \nodata & \nodata & $0.0215 \pm 0.0023$ 

\enddata
\tablecomments{The same notation used as in Table \ref{expansiontable}.}
\tablenotetext{a}{Errors do not include model uncertainties.}
\label{expansiontablea}
\end{deluxetable}

We detect expansion in both the outer and the inner shell (Table
\ref{expansiontablea}). The outer shell expansion rate is $0.0279 \pm
0.0018$\%\ yr$^{-1}$, marginally faster than for the whole shell.
Expansion of the inner shell appears slower by nearly $1/4$, $0.0215
\pm 0.0023$\%\ yr$^{-1}$, but the errors are large. No differences in
expansion rates are expected in the Sedov phase for remnants expanding
either into uniform ambient medium or into the progenitor wind. But
the sharp inner edge seen in \object{G11.2$-$0.3} is not expected in
Sedov models, so any differences in the inner and outer shell
expansion rates are of considerable interest for understanding the
complex dynamics of the remnant.

Expansion of \object{G11.2$-$0.3} at radio wavelengths was determined
by comparison of 2001--2002 and 1984--1985 VLA observations at 20 cm
and 6 cm \citep{tam03}. Our newly-determined expansion rate is in
better agreement with the 6 cm rate of $0.029 \pm 0.010$\%\ yr$^{-1}$
than with $0.042 \pm 0.009$\%\ yr$^{-1}$ measured at 20 cm (errors
quoted here are the rms deviation about the weighted mean of
individual measurements along the remnant's circumference).  As
discussed by \citet{tam03}, it is not clear why the radio expansion
measurements at 20 cm and 6 cm do not match each other better, but
significant and hard to quantify systematic effects might be important
in radio interferometric observations. Expansion rates measured with
{\it Chandra} are less susceptible to systematic errors than the radio
rates, so we consider them more reliable.

At the shell's outer radius of $3.0d_5$ pc (here $d_5$ is the distance
in units of 5 kpc), the mean blast wave speed $v_b$ becomes $810d_5$
km s$^{-1}$. (We assumed here that the blast wave radius
coincides with the shell's outer radius, but there is a possibility
that the blast wave has already propagated farther out into a
low-density ambient medium \citep{chevalier05}. In this case, the blast wave
could have also encountered much denser gas in several directions beyond the
shell, accounting for the presence of outlying H$_2$ filaments
\citep{koo07}. However, the absence of radio or X-ray emission would then be
hard to explain.) The distance to
\object{G11.2$-$0.3} has been estimated from \ion{H}{1} absorption
measurements, ranging from 4.4 kpc \citep[assuming circular Galactic
  rotation;][]{green04} to 5.5 -- 7 kpc \citep[allowing for
  noncircular motions;][]{minter08}. When combined with uncertainties
in the expansion rate measurements, $v_b$ is somewhere within
700--1200 km s$^{-1}$, corresponding to a mean shock temperature
$kT_b$ of $0.6$--$1.7$ keV. This is consistent with X-ray spectra.
\citet{vasisht96} reported an electron temperature of 0.8 keV by
modeling the {\sl ASCA} spectrum of \object{G11.2$-$0.3} with the
nonequilibrium ionization spectral model of \citet{masai84}, while
\citet{roberts03} found a lower electron temperature of 0.6 keV in the
SE using the 2000 {\it Chandra} data in combination with {\tt vpshock}
and {\tt srcut} models in XSPEC \citep{reynolds99,borkowski01}. For
the small regions shown in Figure \ref{shell13}, our spectral fits
result in temperatures ranging from 0.6 keV to 0.8 keV
(Fig.~\ref{spectra13}). We find an intermediate temperature of 0.7 keV
by fitting spectra of the inner and outer shell with an APEC-based
plane shock model plus a power law component to account for the excess
of high-energy photons seen in X-ray spectra. (It is not clear at this
time whether this excess is of thermal or nonthermal origin, but the
temperature of the bulk of X-ray emitting gas depends only weakly on
how this excess is modeled.) The mean shock temperature is expected to
be higher than the electron temperature in young SNRs because of the
preferential heating of ions in fast collisionless shocks, so there is
no obvious conflict between the measured expansion rate and
temperature of the X-ray emitting gas for distances between 4.4 and 7
kpc. If $d = 7$ kpc then very fast ($>1000$ km s$^{-1}$) shocks might 
be present in \object{G11.2$-$0.3}.

\section{Origin of the Ambient Medium and Age of G11.2$-$0.3}

The measured expansion rate of $0.0277 \pm 0.0018$\%\ yr$^{-1}$ allows
us to place constraints on the remnant's age $t_{SNR}$. For a blast
wave radius $r_b$ increasing as $t^m$, $v_b = m r_b/t_{SNR}$, and
$t_{SNR} = m r_b/v_b$. The measured expansion rate is equal to
$v_b/r_b$, so $t_{SNR} = 3600m$ yr.  The upper limit to the remnant's
age is 3600 yr for undecelerated ($m=1$) expansion, but significant
deceleration must have taken place in \object{G11.2$-$0.3} because of
its advanced dynamical age where nearly all of the SN ejecta have been
shocked by now. We use self-similar Sedov models to estimate the
deceleration parameter $m$. For expansion into uniform 
ISM, $m=2/5$ and $t_{SNR}=1400$ yr. This might
be considered as the lower limit to the remnant's age. But
\citet{chevalier05} suggested that the ambient medium in
\object{G11.2$-$0.3} is of circumstellar origin, and was ejected prior
to the SN explosion by its progenitor. For CSM with $\rho \propto
r^{-2}$, produced by steady-state mass loss, $m = 2/3$ and $t_{SNR}$
increases by 5/3 to $2400$ yr. The unknown density distribution of the
ambient medium affects our estimates of the remnant's age to a larger
degree than the modest (at most 20\%) errors in the measured expansion
rates.
 
Single low-mass CC progenitors do not have strong stellar winds and
they do not lose appreciable amounts of mass prior to their
explosions. In this case, after passage through a weak stellar wind
containing very little mass, the blast wave is expected to expand into
the ISM, assumed here to be uniform.  (Alternatively, expansion might
be into a uniform but low density bubble blown by a fast wind during
the main-sequence stage of the progenitor's evolution.)  At the Sedov
stage of evolution, the swept-up ISM mass $M_{\rm sw}$ must be
significantly larger than the ejecta mass. The mass of X-ray emitting
gas in \src\ can be estimated from the X-ray emission measure $EM$
although results are sensitive to $d$ (assumed to be between  
4.4 and 7 kpc).  At $d=4.4$ kpc, 
we obtain $EM = 1.0 \times 10^{59}$ cm$^{-3}$ by fitting
an X-ray Sedov model with the mean shock temperature of $kT_s=0.60$
keV (corresponding to a shock speed of 720 km s$^{-1}$) to the shell
spectrum, and scaling it up by 30\% to account for the missing shell
emission from the remnant's center. (The spectral fits are not
statistically acceptable, but our goal here is just to obtain rough
estimates for $EM$. In the spatially-integrated X-ray Sedov models
available in XSPEC, the amount of collisionless electron heating at
the blast wave, $T_{es}/T_s$, is assumed not to vary with time, but we
considered it as a free parameter in view of our poor knowledge of how
electrons are heated in fast collisionless shocks. Electrons are
subsequently heated downstream of the blast wave through Coulomb
collisions with hot ions, so post-shock electron temperatures $T_e$
depend on the assumed $kT_s$ and fitted $T_{es}/T_s$ and ionization
timescale $\tau$. See \citet{borkowski01} for more details about these
Sedov models.) With $r_b=2.6$ pc, the preshock density $n_0$ is 4
cm$^{-3}$, and the mass of X-ray emitting gas is only 11
$M_\odot$. Since the ejecta mass $M_{ej}$ must be more than 6
$M_\odot$ even for the least massive ($\sim 8 M_\odot$) CC progenitor,
$M_{\rm sw} < 5 M_\odot < M_{ej}$, and the remnant cannot be in the
Sedov stage of the evolution.  In disagreement with observations, the
reverse shock is unlikely to propagate all the way to the center for
such a dynamically young remnant.  Somewhat better agreement arises if
\object{G11.2$-$0.3} were much farther away than 4.4 kpc. From X-ray
Sedov model fits with $kT_s=1.18 \left( v_b/1000\ {\rm km
  s}^{-1}\right)^2$ keV, we find that $EM$ is nearly independent of
the assumed distance. (When model parameters do not depend on $d$, $EM
\propto d^2$. But here $kT_s \propto v_b^2 \propto d^2$, and the
increased X-ray emissivity of the Sedov model at higher blast wave
speeds leads to an approximately constant $EM$.) This implies that
$n_0 \propto d^{-3/2}$ and $M \propto d^{3/2}$, so $n_0 = 3.4
d_5^{-3/2}$ cm$^{-3}$ and $M = 13d_5^{3/2}$ $M_\odot$.  At 7 kpc, $n_0
= 2$ cm$^{-3}$ and the total mass of the X-ray emitting gas is about
22 $M_\odot$. The ejecta mass $M_{ej} > 6M_\odot$ comprises of at
least $\sim 0.3$ of this amount, so the remnant might still not be
fully in the Sedov evolutionary stage.

Progenitors significantly more massive than $8 M_\odot$ can lose the
10--20 $M_\odot$ present in \object{G11.2$-$0.3}. For a single
progenitor, this might occur through strong stellar winds, while in
close binaries mass loss is further enhanced through tidal
interactions. In the Sedov solution with a steady-state stellar wind
with $\rho = D r^{-2}$ (or $n_0 = D_n r^{-2}$, where $D_n=D\left(\mu
m_p\right)^{-1}$, $\mu = 1.4$ is the mean mass per hydrogen atom in
atomic mass units, and $m_p$ is the proton mass), the emission measure
$EM$ is equal to $\frac{n_e}{n_H} \frac{64\pi D_n^2}{5r_b}$ (for
cosmic abundances $n_e/n_H=1.23$). Just as for a uniform ambient
medium, the mass of X-ray emitting gas and the ambient density (i.e.,
$D$ or $D_n$) can be found by estimating $EM$ and $r_b$. Since an
X-ray Sedov wind model is not available in XSPEC, we assumed
plane-parallel geometry but unequal electron and ion temperatures and
used the plane shock models {\tt npshock} and {\tt vnpshock} in XSPEC
in our estimates of $EM$ (the X-ray fitting procedure is the same as
described above).  The estimated emission measures are again around $1
\times 10^{59}$ cm$^{-3}$, but the preshock densities are about half
as large ($n_0 = 2$ cm$^{-3}$ at 4.4 kpc, and $n_0 =0.9$ cm$^{-3}$ at
7 kpc) and masses are 50\%\ higher (15--30 $M_\odot$) than for the
uniform ambient medium. These lower density and higher mass estimates
are caused by differences in the model postshock density distributions
(density drops more slowly with decreasing radius in the Sedov wind
model than in the standard Sedov model). The wind strength is $D_*
\sim 3$ (where $D_*=D/1.0 \times 10^{14}$ g cm$^{-1}$), corresponding
to $\dot{M} \sim 10^{-4} M_\odot$ yr$^{-1}$ and $v_w=15$ km
s$^{-1}$. This is at the upper range of mass-loss rates inferred for
Type IIL and IIb SNe \citep[][]{chevalier09,smith14}.

The post-shock density $\rho$ increases linearly with radius in the
Sedov wind model.  This results in an X-ray shell with the surface
brightness profile of $2^{-1/2} \left( 1 - (r/r_b)^2 \right)^{1/2}
\left( 2(r/r_b)^2 + 1 \right)$ (we assumed here that the X-ray
emissivity scales as $\rho^2$, and normalized the brightness to a
peak of 1). The maximum surface brightness peaks at a rather small ($r =
2^{-1/2}r_b$) radius, and the shell is not very distinct (the central
surface brightness is lower than the peak by only $2^{1/2}$). Such
a thick and indistinct shell is clearly inconsistent with the
observations, since the X-ray shell in \object{G11.2$-$0.3} is quite
prominent and not overly broad. The density must drop quite rapidly
toward the center, as evidenced by the sharp inner shell
edge at several locations within the remnant. This means that the
shocked masses have been overestimated, and preshock densities
underestimated when using the Sedov wind model. A wind-like ($n_0 =
D_n r^{-2}$) density distribution for the CSM appears unlikely even if
one allows for departures from the Sedov wind model. (This
self-similar model is only asymptotically valid in the limit when the
shocked ejecta mass is negligible compared to the shocked ambient
mass, while young remnants such as \object{G11.2$-$0.3} might still be
at earlier evolutionary stages).  In order to keep $EM$ fixed while
making the shell more prominent than in the Sedov wind model, a
decrease in the swept-up mass and an increase in the preshock density
would have been required. These two requirements are mutually
exclusive under a steady-state mass loss hypothesis.  Note that these
estimates of preshock densities and shocked CSM masses rely on the
assumption that most of the X-ray emission is produced by the swept-up
CSM. This assumption is justified given the lack of evidence for
enrichment in heavy elements anywhere within the remnant, even in its
interior.
The
shocked ejecta contribution to X-rays might be nonnegligible, but very
likely it is not dominant. The CSM in \object{G11.2$-$0.3}, if
present, is likely to have a more complex density distribution than
described by a steady-state stellar wind.

A CSM origin for the swept-up gas in \object{G11.2$-$0.3} still
remains viable although its ambient density distribution is not well
described by a steady-state stellar wind. The absence of a sharp and
well-defined outer boundary of the remnant might result if the blast
wave had already overrun the slow wind of the SN progenitor. But this
will not make the X-ray emitting shell more prominent than in the
Sedov wind model, and aside from a few isolated outer knots, no
widespread X-ray or radio emission is seen beyond the outer shell
boundary. A subenergetic explosion might also be required since the
shell velocity is rather low. Alternatively, the progenitor's wind
might have been less dense at the very final stages of its evolution
prior to the SN, with a significant fraction of mass lost $\sim 10^5$
yr before the explosion through a dense and slow outflow. The wind
density is generally expected to drop with time when a single massive
progenitor moves from red to blue across the HR diagram because of
heavy mass loss. The mass-loss history becomes more complex and less
understood in close binaries, but steady-state mass loss becomes even
less likely. If the progenitor loses enough material to expose its
helium core (or nearly expose it with a residual hydrogen envelope
still present), a radiatively-driven fast wind is expected to sweep
material ejected through slow and dense winds in the prior
evolutionary stages into dense shells. The blast wave in
\object{G11.2$-$0.3} might be now moving through undisturbed CSM lost
$\sim 10^5$ yr ago when the mass loss rate was much higher than
immediately prior to the explosion. Alternatively, the blast wave
might be propagating through a dense swept-up shell or it might have
already exited into much more tenuous gas located beyond its outer
boundary. In either case, the remnant's dynamics becomes complex, and
its understanding requires reliance on hydrodynamical
simulations. Within the framework of such future investigations, the
measured expansion of the shell is expected to provide refined
constraints on the remnant's age and SN explosion properties.

The strongest evidence for a CSM (as opposed to ISM) origin of the
swept-up ambient medium is the presence of strongly asymmetric soft
thermal emission in the interior of \object{G11.2$-$0.3}.
Its
asymmetric morphology argues strongly against an ISM origin, as this
requires a highly improbable density distribution centered around the
explosion site. Strongly asymmetric mass loss provides a natural
explanation for this interior emission. Such interior emission might
become prominent in SNR 1987A once its blast wave envelops its famous
bright inner ring, if this ring marks the inner edge of a much larger
torus.  Another example is provided by the dusty torus in the red
supergiant (RSG) WOH G64 in the Large Magellanic Cloud
\citep{ohnaka08}. With an initial main-sequence mass of 25 $M_\odot$
and with a total ejected mass of 3--9 $M_\odot$, this RSG has already
lost a significant fraction of its mass through a highly asymmetric
outflow.  Such a massive and asymmetric outflow is expected in close
binary systems because of tidal effects, although firm evidence for
the presence of a binary companion in WOH G64 is still lacking
\citep{levesque09}.  Rotation might also result in a strongly
asymmetric outflow even for a single SN progenitor, and both rapid
rotation \citep{chita08} and binarity \citep{podsiadlowski91} have
been invoked as the origin of the observed CSM asymmetry in SNR 1987A.

Extreme asymmetry is expected if most mass was lost by the progenitor
when it was undergoing Roche-lobe overflow. A possible detection of
a close binary companion to SN 2011dh by \citet{folatelli14} suggests
that this is how mass was lost by the progenitor of this nearby Type
IIb SN. According to the binary evolution model of
\citet{benvenuto13}, confirmed by detection of the companion, the
progenitor underwent several episodes of high mass-transfer Roche-lobe
overflow. The most recent transfer took place $\sim 10^5$ yr prior to
the explosion, so massive and asymmetric CSM is expected to be present
in SN 2011dh at pc-scale distances from the explosion site.

Systematic theoretical investigations of effects that strongly
asymmetric outflows imprint on both SNe and SNRs are
lacking. Exploratory 2D hydrodynamical simulations with a strongly
asymmetric wind in the context of a Type Ia explosion indicate that
the interior emission seen in Kepler's SNR can be explained in this
framework \citep{burkey13}. Like Kepler's SNR for Type Ia explosions,
\object{G11.2$-$0.3} becomes important for understanding the origin of
asymmetric mass loss in CC SNe. The pulsar wind at its center probes
the interior of the remnant, while the shocked thermal gas and IR
emission provide information about both the shocked ambient gas and
the shocked ejecta. There is much to learn by studying
\object{G11.2$-$0.3} in more detail, but it becomes obvious that
multidimensional hydrodynamical simulations are necessary to make
further progress. Although \citet{kaplan06} found no IR counterpart to
the pulsar that might be expected in the presence of a surviving
binary companion bound to the pulsar, the binary could have been
disrupted by the explosion.  Just as for SNR 1987A, the origin of the
CSM asymmetry remains unclear at this time.

\section{Pulsar-Wind Nebula Analysis}

The confirmation of a young age for \src, even though it cannot be
the result of an event in CE 386, supports the conclusion that the
pulsar was born at nearly its present period.  For a constant braking
index $n$, the true age of the pulsar is given by
\begin{equation}
t = \frac{P} {(n-1) {\dot P}} 
  \left[1 - \left(\frac{P_0} {P} \right)^{n-1}\right].
\end{equation}
Taking a typical braking index of 2.5 and an age estimate of 2000 yr,
we find $P_0 = 0.96 P = 62$ ms, consistent with the estimate of
\cite{torii99}, and implying an initial energy $I \Omega_0^2/2 =
5\times 10^{48}$ erg, for a moment of inertia of $10^{45}$ g cm$^2$.
Spindown with constant magnetic field and braking index also gives the
time-dependence of pulsar luminosity by
\begin{equation}
L = \frac{L_0}{\left(1 + t/\tau \right)^{ \frac{n + 1} {n - 1} }}
\end{equation}
with the slowing-down timescale $\tau$ given by
\begin{equation}
\tau = \frac{2}{n - 1} {P \over {\dot P}} - t \cong 60,000 \ {\rm yr}.
\end{equation}
That is, the power input into the PWN from the pulsar has been
roughly constant since the supernova.

We can attempt to interpret the brightness ratio between the NE jet
and the bright knots in the SW jet as Doppler boosting.  For
synchrotron emission with photon index $\Gamma$, the jet/counterjet
brightness ratio, for jets making an angle $\theta$ with the line of
sight and traveling at speed $v = \beta c$, is
\begin{equation}
\frac{I({\rm jet})}{I({\rm counterjet})} =
  \left( \frac{1 + \beta \cos\theta} {1 - \beta \cos\theta} \right)^a
\end{equation}
where $a = \Gamma + 1$ for a continuous jet and $a = \Gamma + 2$ for
discrete blobs \citep{lind85}.  We estimate $\cos\theta = 0.5$ from
the torus aspect ratio, and a brightness ratio of 3.  Using the
expression for a continuous jet, and taking the overall PWN photon
index of 1.78, we obtain $\beta = 0.4$.  Using the knot value of
$\Gamma = 1.36$, we find $\beta = 0.46$; and for the discrete blob
exponent (with $\Gamma = 1.36$), we obtain $\beta = 0.32.$ Thus mildly
relativistic flow could produce the X-ray brightness ratio we see, 
as well as account for a similar brightness asymmetry seen at radio
wavelengths.
These
values of $\beta$ are quite typical for pulsar torus/jet models
\citep{ng04}.  For $\beta = 0.4$, the sky-plane speed would be $5.1''$
yr$^{-1}$, thus requiring about 8 yr to reach the ends of the jet.
From standard synchrotron expressions, a magnetic field of less than
about 140 $\mu$G would result in a synchrotron-loss time of 8 yr or
greater for electrons radiating 8 keV photons.  However, the flow
would need to decelerate rapidly downstream of the bright knots, as
the jet does not appear to expand (barring some peculiar behavior of
magnetic field, such as its rapid disappearance).  Rapid deceleration
is in fact a consequence of the spherical MHD model of
\cite{kennel84}, though the geometry here is clearly more complex.
Some kind of magnetic braking may be involved \citep{komissarov04}, as
seems to be required in other cases as well.  In any case, we can
rule out that the knots are moving with anything like $\beta = 0.4$;
they would need to be stationary structures such as internal shocks,
through which fluid passes.

We point out that while extrapolations of the PWN X-ray power-law
spectrum back to the radio result in a ``break'' frequency of 8 GHz
with a change in slope of about 0.5, the expected value for
synchrotron losses in a homogeneous source with constant electron
injection, this feature in \src\ is unlikely to be due to losses, as
it would require a magnetic field strength of about 3 mG for an
assumed remnant age of 2000 yr.  This value would imply a loss time for
8 keV-emitting electrons less than the 8 yr estimated above by
$(3/0.14)^{-1.5}$, or about a month, so that the jets could not
possibly extend as far as they are observed without clear spectral
steepening.  

In our interpretation, the PWN has already been compressed by the
return of the reverse shock.  However, unlike most other PWNe, the
much longer pulsar slowdown timescale for \src\ means that the pulsar
energy input is about the same as its initial value, rather than far
weaker, which is more typical.  We interpret the more extended radio
and X-ray nebulae as the remnants of the pre-reverse-shock PWN, and
the X-ray jets as more recently produced features.  The integrated
spectrum can then be quite complex \citep{reynolds84}, with multiple
bends and breaks between radio and X-rays, even without invoking
intrinsic spectral structure in the electron distribution produced at
the original pulsar-wind termination shock.

\cite{chevalier05} estimates a minimum nonthermal energy for the PWN
in \src\ which implies a minimum pressure of about $10^{-10}$ dyn
cm$^{-2}$.  While he approximates the PWN as a sphere, the estimate is
unlikely to be dramatically in error.  It implies a magnetic field of
order 50 $\mu$G, implying an energy loss timescale of about 36 yr for
8 keV-emitting electrons.  This in turn requires a mean flow velocity
of about 20,000 km s$^{-1}$ to avoid producing spectral steepening at
the jet ends -- less than our inferred transrelativistic speeds for
Doppler boosting in the knots, but inconsistent with the absence of
outward motion of jet ends.  This minimum pressure is, however, about
two orders of magnitude below the pressure to be expected in the
remnant interior if all the SN energy has been thermalized (see
below).  We discuss this discrepancy in the following section.

\section{Discussion and Conclusions}

The high visual extinction toward \src, obtained both from [Fe II]
observations \citep{lee13} and from our absorption measurements toward
the PWN and the shell, definitively rules out the association of
\src\ with any naked-eye event seen on Earth, in particular with the
CE 386 event, unfortunately removing \src\ from the short list of SNRs
with known ages.  However, our mean shell expansion rate of $0.0277
\pm 0.0018$\% yr$^{-1}$ gives an age in a comparable range, for a
plausible range of expansion parameter $m$ between 0.4 and 2/3, of
1400 -- 2400 yr -- comparable to the 1629 yr resulting from the CE 386
association.  We confirm that \src\ is one of the three or four
youngest shell CC remnants in the Galaxy -- perhaps third, behind Cas
A and Kes 75 (containing a young pulsar, with spindown age $\lapprox$
1700 yr; Mereghetti et al.~2002, Livingstone et al.~2006).

In the PWN, we have identified a torus-like structure in radio and
X-rays. Such structures are commonly seen in PWNe, but only in X-rays;
a sharply defined radio torus $+$ jet is unique to \src.
Furthermore, the jet in \src\ completely
dominates the torus -- in fact, the bulk of the X-ray PWN emission is
from the jet, unlike most PWNe with jet/torus structure.  We attribute
this characteristic to the very long pulsar slowdown timescale, so
that the pulsar energy input is almost the same as at birth.  Mildly
relativistic outflow could explain the brightness asymmetry between SW
and NE jets, though this interpretation is not without problems.

Our X-ray spatial analysis shows several anomalous features for a
shell SNR.  The presence of significant asymmetric interior emission
requires significant departures from the simple self-similar driven
wave or Sedov evolutionary phases.  The sharp inner edge of the outer
shell and the hard X-ray emission found near this edge are
unexplained in any simple picture.  No sharp X-ray rims are
found, unlike all young Type Ia remnants, where they are found to have
nonthermal spectra and indicate significant magnetic-field
amplification \citep{parizot06}.  (Some CC remnants also show these
rims, such as Cas A and the X-ray-synchrotron-dominated remnants
G347.3$-$0.5 (RX J1713.7$-$3946) and G266.2$-$1.2 (RX J0852.0$-$4622).
See \cite{reynolds12} for discussion and references.)

The X-ray spectrum of the shell can be reasonably well described by a
thermal, plane-shock model with temperatures of 0.6 -- 0.8 keV,
consistent with inferred shock velocities ranging from 700 km s$^{-1}$ 
\citep[at 4.4 kpc;][]{green04} to 1100 km s$^{-1}$
\citep[at the largest allowed distance of 7 kpc;][]{minter08} that are
required to account for the measured expansion of the shell.
The emission measures from fits with a more-sophisticated X-ray Sedov model
indicate a mass of X-ray emitting gas of
about $M = 13 d_5^{3/2}$ $M_{\odot}$, too low for the
remnant to be fully in the Sedov stage of evolution into a uniform
medium.  The bar of central emission running at roughly position
angles (100$^\circ$ -- 280$^\circ$) has a thermal spectrum, with no
clear evidence for enhanced abundances.  The presence of thermal
emission near the center of the remnant, and clear morphological
correspondences between the inner thermal emission and the PWN,
indicate that the reverse shock has already returned to the center and
re-shocked all the SN ejecta.  Since it is expected
\citep[e.g.,][]{chevalier05} that \src\ is expanding into a stellar
wind, a reanalysis of the spectrum for a wind model increases
the inferred shocked mass to 15 -- 30 $M_\odot$.  However, the
predicted morphology of a remnant running into a uniform $r^{-2}$
density profile is not consistent with the sharp shell we observe.
The assumption of a smooth, spherically symmetric wind CSM is
apparently far too simple to explain what we see in \src.

We summarize here the scenario we propose for this complex object.  We
adopt Chevalier's (2005) classification of \src\ as resulting from a
category of supernovae exploding after losing most of their
envelopes (several solar masses) prior to explosion.  Such substantial
mass loss is likely to be quite asymmetric, perhaps due to a binary
companion, and we invoke the possibility of a disk wind.

After the explosion, the supernova blast wave has encountered this
lost mass, both radially and azimuthally inhomogeneous, and has by now
swept up of order 10 solar masses.  The density currently just ahead
of the blast wave is quite low, resulting in a diffuse rather than
sharp outer edge of shell emission.  The reverse shock has moved all
the way back to the center, reheating all ejecta, but CSM
inhomogeneities have left characteristic imprints, such as the sharp
inner edges of the outer shell visible in some locations and the
limb-brightened hard X-ray emission in the remnant's interior.  We
interpret the presence of hard X-rays near the inner edge of the
remnant's outer shell as evidence for an abrupt change in the density
distribution of the ambient CSM.  This is consistent with the sharp
inner edge of the shell.  The shocked low-density gas in the interior
of the remnant, including the bulk of the SN ejecta, does not radiate
as efficiently as the much denser material in the outer shell, so it
is difficult to detect its faint thermal X-ray emission against the
bright shell. Only the most dense shocked ambient gas is visible in
thermal X-rays as the anomalous bar of interior emission. X-ray
emission from the more tenuous gas is clearly seen only at high photon
energies where gas in the shell radiates less efficiently because of
its low (0.6 -- 0.8 keV) temperature.  This emission might be of
nonthermal origin.  (Electrons radiating 8-keV synchrotron photons
can survive for $\sim 1000$ yr if $B \lapprox 5\ \mu$G, as might be true in
the low-density interior.  Internal diffuse X-ray synchrotron emission is also
seen in the comparably-aged remnant RCW 86 [Williams et al.~2011], a
remnant thought to result from a cavity explosion.) 
Shocked ejecta have been unambiguously detected
only in
some very dense clumps which have cooled and are the source of the observed
[Fe II] near-IR emission.  The pulsar was born with a very low velocity
and is still near the remnant center.  The PWN it produced has been
compressed by the return of the reverse shock, but in a fairly
symmetrical fashion.  The pulsar continues to inject relativistic
fluid into the PWN, in a jet/torus configuration in which the jets,
rather than the torus, are the most apparent.

This scenario can accommodate most of what is known about \src, but
leaves many important questions unanswered.  Most prominent is the
nature of the mass lost: evidently its spatial distribution is uneven
in all respects. The inferred density discontinuity in the CSM density
distribution points to a major change in the nature of mass loss after
ejection of the bulk of the stellar envelope in slow and dense outflows.
It is possible that the
progenitor was a compact blue star at the time of the explosion,
with only a residual hydrogen envelope left on top of its nearly
stripped helium core. This would have resulted 
in a (compact) cIIb SN \citep{chevsod10} instead of a
IIL or an (extended) eIIb SN expected for much larger red or yellow
supergiant progenitors that suffered substantial but
not so extreme mass loss.
Alternatively, the entire H envelope might have
been stripped, and the SN was of Type Ibc. Most 
Type cIIb/Ibc SNe arise from explosions of tidally-stripped binary progenitors
with main-sequence masses less than $\sim 25 M_\odot$ rather than from 
explosions of the classical Wolf-Rayet stars originating in more massive stars
\citep[][]{smith14}.
The progenitor of G11.2$-$0.3 was
likely a low-mass He star in a close binary system, possibly with a residual
H envelope still present as inferred and modeled for the Type cIIb SN 2008ax
\citep{chevsod10,folatelli15}. 
Such compact progenitors of Type cIIb/Ibc SNe lose mass in
fast stellar winds that compress denser and more slowly moving
material ejected in the prior evolutionary stages into dense shells, with
the low-density shocked fast wind filling the interiors of these shells.
Their strong ultraviolet radiation heats dust that is present in these
shells. Progenitor systems similar to what we infer for G11.2$-$0.3, consisting 
of a low-mass He star, a binary companion that is necessary to strip the
H envelope, and a compact IR-emitting shell,
are likely present among many compact circumstellar bubbles discovered
by {\sl Spitzer} that are powered by massive stars
\citep{wachter10,gvamaradze10,flagey14}.

The nature of interior material is, however, still debatable.  While
CSM in a disk is a likely arrangement for strongly asymmetric dense
CSM swept up by the fast wind of a stripped-envelope progenitor, it is
also possible that this material is highly asymmetric ejecta.  While
spectral analysis does not show obvious heavy-element enhancements
anywhere in \src, a mixture of shocked CSM and ejecta might not show
clear spectral signatures, particularly if the ejecta were much cooler
than the CSM.  A mechanism for producing such asymmetric ejecta is
unknown at this time.

The thermal pressure in the outer shell is several
$\times 10^{-8}$ dyn cm$^{-2}$,
and we expect a comparable central pressure if all the SN
energy has been thermalized.
The estimate of a central thermal pressure of at least $\sim 10^{-8}$
dyn cm$^{-2}$, which results from the assumption that only mild pressure
variations are present within the interior of G11.2$-$0.3,
causes serious difficulties in interpreting the PWN.  We showed above that
the minimum (equipartition) nonthermal
pressure in the PWN is two orders of magnitude less.  Now there is no
guarantee that equipartition must hold.  However, if the magnetic
energy dominates, bringing the PWN pressure up to the central thermal pressure
would require a magnetic-field strength of about 0.6 mG, in which
synchrotron lifetimes of 8 keV-emitting electrons would be only one
year -- clearly at odds with the lack of spectral steepening in the
jets (which are 2 -- 3 lt-yr long).  The maximum magnetic-field
strength consistent with that constraint is about 0.2 mG, providing
only one-tenth the required pressure.  If the PWN is
particle-dominated, not an impossibility, the magnetic-field strength
is basically unconstrained.  3-D relativistic MHD simulations
\citep{porth14} show that magnetic dissipation can lower the magnetic
energy to a small fraction of the particle energy.  These simulations
also show that a young PWN is a complex object, with strong departures
from symmetry in magnetic-field strength and (to some extent) in
pressure, so all conclusions drawn from assumptions of spherical
symmetry should be regarded as provisional.

The high external pressure also implies a pulsar-wind termination 
shock radius given by
\begin{equation}
r_s = \left( \frac{\dot E}{4 \pi c P_{\rm ext}} \right)^{1/2} 
  \sim 0.01 \ {\rm pc}
\end{equation}
or about $0.5''$, far smaller than the size of the torus -- but
given the drastic oversimplifications (mainly spherical symmetry) under
which this simple estimate is obtained, perhaps this is not a concern.
In any case, the inner edge of the torus, where the shock would be
located, could well be at a much smaller radius.

The inferred thermal pressure, while high, implies a low explosion
energy for a distance of 4.4 kpc: $E_{\rm SN} \sim 2 \times 10^{50}$
erg. (We accounted for the kinetic energy of the shell in our estimates of the
explosion energy but it is a minor (20\%--25\%) contribution to the total
energy budget.) This estimate scales with the volume, or more accurately,
$E_{\rm SN} = 2.5 \times 10^{50} d_5^{7/2}$ erg; at 7 kpc it rises by about
a factor of 4, more consistent with a typical CC explosion.  It
is possible that the outer shell we observe is not the outermost
extremity of the SNR.  In addition to a few obvious knots at larger radii,
it is possible that hot thermal material is able to
``leak'' through the shell in spots, where if the density is
sufficiently low it might not be readily detectable in X-rays or
infrared.
However, it would
be more difficult to hide such material from a deep radio image, so it
would be worthwhile to revisit \src\ with the greatly enhanced
capabilities of the JVLA to (among other goals) search for such a
halo. If present, this halo might also shed light on the origin of outlying
H$_2$ filaments found by \citet{koo07}. It is possible that these filaments
mark locations where the blast wave has encountered much denser than average
material after breaking through the shell. 

The ejecta of a stripped-envelope SN contain little (if any) H, while
dense CSM swept up by the blast wave consists mostly of H. The shocked
CSM and the shocked ejecta in G11.2$-$0.3 can be distinguished by
the presence or absence of H lines in their near-IR spectra and by their
kinematics. \citet{koo07} detected the Br$\gamma$ line in a spectrum of
the brightest emission filament in the SE, and H$_2$ emission lines in
a spectrum of a spatially-localized region in the S, so the emission there
comes mostly from shocked CSM. Highly blueshifted ($\sim 1000$ km
s$^{-1}$) knots near the center are devoid of H \citep{moon09},
consistent with ejecta of a stripped-envelope SN. The shocked, H-rich
CSM has been accelerated by the blast wave, but its velocity is
expected to be at most one or two hundred km s$^{-1}$, far less than
the $\sim 1000$ km s$^{-1}$ found for the H-poor ejecta. \citet{moon09}
proposed that the ejecta are rich in Fe, but other elements such as He
might still be dominant. (The upper limit to the strength of the
\ion{He}{1} 1.083 $\mu$m line cited by \citet{moon09} is not
particularly constraining.) Deeper spectra are needed in order to
learn more about the composition of the IR-emitting ejecta, but we
predict that little (if any) H will be found within the SN ejecta.

Further understanding of this interesting object will require
multi-dimensional hydrodynamic simulations, which can test some of the
features of our proposed scenario.  Additional observational efforts
in radio and in near-IR are the most likely to cast additional light.
A search for a possible surviving binary companion would be quite
challenging but, if successful, would be a major advance.
\src\ serves as additional evidence, if any were needed, that
supernovae, and their remnants, are fundamentally three-dimensional
objects, and their full understanding will require three-dimensional
theory.

\acknowledgments
This work was supported by NASA through the {\sl Chandra} General Observer
Program grant GO3-14076A.
We are grateful to Cindy Tam for reduction of the VLA data. We acknowledge
help from Andrew Moseby in reprocessing and preparing {\sl Chandra} data
of G11.2$-$0.3 for analysis. The scientific results reported in this article
are based on observations made by the {\sl Chandra} X-ray Observatory, and
by the {\sl Karl G. Jansky} Very Large Array at the National Radio Astronomy
Observatory. 
The National Radio Astronomy Observatory is a facility of the National Science
Foundation operated under cooperative agreement by Associated
Universities, Inc. This research has made use of software provided by the
{\sl Chandra} X-ray Center (CXC) in the applications packages {\sl CIAO} and
{\sl ChIPS}. We acknowledge use of various open-source software 
packages for Python, including Numpy, Scipy, Matplotlib, Astropy
(a community-developed core Python package for Astronomy), and
APLpy\footnote{APLpy is an open-source plotting package for Python hosted at
http://aplpy.github.com}.


\begin{thebibliography}{}

\bibitem[Andersen et al.(2011)]{andersen11}
Andersen, M., Rho, J., Reach, W. T., Hewitt, J. W., \& Bernard, J. P.
2011, ApJ, 742, 7

\bibitem[Arendt(1989)]{arendt89}
Arendt, R. G.
1989, ApJS, 70, 181

\bibitem[Arnaud(1996)]{arnaud96}
Arnaud, K. A. 1996, in Astronomical Data Analysis and Systems V, 
eds. G.~Jacoby \& J.~Barnes, ASP Conf.~Series, v.101, 17

\bibitem[Becker et al.(1985)]{becker85}
Becker, R. H., Markert, T., \& Donahue, M. 1985, ApJ, 296, 461

\bibitem[Benvenuto et al.(2013)]{benvenuto13}
Benvenuto, O. G., Bersten, M. C., \& Nomoto, K. 
2013, ApJ, 762, 74

\bibitem[Borkowski et al.(2001)]{borkowski01}
Borkowski, K. J., Lyerly, W. J., \&\ Reynolds, S. P.
2001, ApJ, 548, 820

\bibitem[Burkey et al.(2013)]{burkey13}
Burkey, M., Reynolds, S. P., Borkowski, K. J., \& Blondin, J. M.
2013, ApJ, 764:63

\bibitem[Carlton et al.(2011)]{carlton11}
Carlton, A. K., Borkowski, K. J., Reynolds, S. P., et al. 
2011, ApJ, 2011, 737:L22

\bibitem[Cash(1979)]{cash79}
Cash, W. 1979, ApJ, 228, 939

\bibitem[Chevalier(2005)]{chevalier05}
Chevalier, R. A. 2005, ApJ, 619, 839

\bibitem[Chevalier(2009)]{chevalier09}
Chevalier, R. A. 2009, in Proc.~Space Telescope Science Institute Symposium 
Series No.~20,
Eds. M.~Livio 
\& E.~Villaver (Cambridge University Press), 199

\bibitem[Chevalier \& Soderberg(2010)]{chevsod10}
  Chevalier, R. A., \& Soderberg, A. M.
  2010, ApJ, 711, L40

\bibitem[Chi\c{t}\u{a} et al.(2008)]{chita08}
  Chi\c{t}\u{a}, S. M., Langer, N., van Marle, A. J., Garc\'{i}a-Segura, G., 
  \& Heger, A.
  2008, A\&A, 488, L37

\bibitem[Cornwell et al.(1999)]{cbb99}
  Cornwell, T., Braun, R.,  \& Briggs, D. S.
  1999, Synthesis Imaging in Radio Astronomy II, 180, 151

\bibitem[Downes(1984)]{downes84}
Downes, A. 1984, MNRAS, 210, 845

\bibitem[Flagey et al.(2014)]{flagey14}
    Flagey, N., Noriega-Crespo, A., Petric, A., \& Geballe, T. R.
    2014, AJ, 148, 34

\bibitem[Folatelli et al.(2014)]{folatelli14}
Folatelli, G., Bersten, M. C., Benvenuto, O. G., et al.
2014, ApJ, 793:L22

\bibitem[Folatelli et al.(2015)]{folatelli15}
Folatelli, G., Bersten, M. C., Kuncarayakti, H., et al.
2015, ApJ, 811:147

\bibitem[Foster et al.(2012)]{foster12}
  Foster, A. R., Ji, L., Smith, R. K., \& Brickhouse, N. S.
  2012, ApJ, 756, 128

\bibitem[Froebrich et al.(2015)]{froebrich15}
  Froebrich, D., Makin, S. V., Davis, C. J., et al.
  2015, MNRAS, 454, 2586
  
\bibitem[Giannini et al.(2015)]{giannini15}
  Giannini, T., Antoniucci, S., Nisini, B., et al.
  2015, ApJ, 798:33

\bibitem[Green(2004)]{green04}
Green, D. A. 2004, BASI, 32, 335

\bibitem[Green(2011)]{green11}
Green, D. A.
2011, BASI, 39, 289 

\bibitem[Green et al.(1988)]{green88}
Green, D. A., Gull, S. F., Tan, S. M., \& Simon, A. J. B.
1988, MNRAS, 231, 735

\bibitem[Grevesse \& Sauval(1998)]{grsa98}
Grevesse, N., \& Sauval, A. J.
1998, Space Sci.~Rev., 85, 1

\bibitem[Gvaramadze et al.(2010)]{gvamaradze10}
  Gvaramadze, V. V., Kniazev, A. Y., \& Fabrika, S.
  2010, MNRAS, 405, 1047

\bibitem[Hester(2008)]{hester08}
Hester, J. J. 2008, ARA\&A, 46, 127

\bibitem[Kaplan \& Moon(2006)]{kaplan06}
Kaplan, D. L., \& Moon. D.-S.
2006, ApJ, 644, 1056

\bibitem[Kaspi et al.(2001)]{kaspi01}
  Kaspi, V. M., Roberts, M. S. E., Vasisht, G., et al.
  2001, ApJ, 560, 371

\bibitem[Kennel \& Coroniti(1984)]{kennel84}
Kennel, C. F., \& Coroniti, F. V.
1984, ApJ, 283, 694

\bibitem[Kilpatrick et al.(2016)]{kilpatrick16}
  Kilpatrick, C. D., Bieging, J. H., \& Rieke, G. H.
  2016, ApJ, 816:1 

\bibitem[Komissarov \& Lyubarsky(2004)]{komissarov04}
Komissarov, S. S., \& Lyubarsky, Y. E.
2004, MNRAS, 349, 779

\bibitem[Koo \& Lee(2015)]{koolee15}
  Koo, B.-C., \& Lee, Y.-H.
  2015, PKAS, in Proc.~12th Asia-Pacific Regional IAU Meeting  
  (arXiv:1502.00048)

\bibitem[Koo et al.(2007)]{koo07}
  Koo, B.-C., Moon, D.-S., Lee, H.-G., et al.
  2007, ApJ, 657, 308

\bibitem[Kothes \& Reich(2001)]{kothes01}
Kothes, R., \& Reich, W.
2001, A\&A, 372, 607

\bibitem[Krishnamurthy et al.(2010)]{krishnamurthy10}
  Krishnamurthy, K., Raginsky, M., \& Willett, R.
  2010, SIAM J. Imaging Sci., 3, 619

\bibitem[Lee et al.(2013)]{lee13}
  Lee, H.-G., Moon, D.-S., Koo, B.-C., et al.
  2013, ApJ, 770, 143

\bibitem[Levesque et al.(2009)]{levesque09}
Levesque, E. M., Massey, P., Plez, B, \& Olsen, K. A. G.
2009, AJ, 137, 4744

\bibitem[Lind \& Blandford(1985)]{lind85}
Lind, K., \& Blandford, R. D.
1985, ApJ, 295, 358

\bibitem[Livingstone et al.(2006)]{livingstone06}
Livingstone, M. A., Kaspi, V. M., Gotthelf, E. V., \& Kuiper, L.
2006, ApJ, 647, 1286

\bibitem[Masai(1984)]{masai84}
Masai, K., 
1984, J.~Quant.~Spectrosc.~Rad.~Transf., 51, 211

\bibitem[Mereghetti et al.(2002)]{mereghetti02}
Mereghetti, S., Bandiera, R., Bocchino, F., \& Israel, G. L.
2002, ApJ, 574, 873

\bibitem[Minter et al.(2008)]{minter08}
Minter, A. H., Camilo, F., Ransom, S. M., Halpern, J. P., \& Zimmerman, N.
2008, ApJ, 676, 1189

\bibitem[Moon et al.(2009)]{moon09}
  Moon, D.-S., Koo, B.-C., Lee, H.-G., et al.
  2009, ApJ, 703, L81

\bibitem[Ng \& Romani(2004)]{ng04}
Ng, C.-Y., \& Romani, R. W.
2004, ApJ, 601, 479

\bibitem[Ohnaka et al.(2008)]{ohnaka08}
Ohnaka, K., Driebe, T., Hofmann, K.-H., Weigelt, G., \& Wittkowski, M. 
2008, A\&A, 484, 371

\bibitem[Parizot et al.(2006)]{parizot06}
Parizot, E., Marcowith, A., Ballet, J., \& Gallant, Y. A.
2006, A\&A, 453, 387

\bibitem[Pinheiro Gon\c{c}alves et al.(2011)]{pinheiro11}
  Pinheiro Gon\c{c}alves, D., Noriega-Crespo, A., Paladini, R., Martin, P. G.,
  \& Carey, S. J.
2011, AJ, 142, 47

\bibitem[Podsiadlowski et al.(1991)]{podsiadlowski91}
  Podsiadlowski, P., Fabian, A. C., \& Stevens, I. R.
  1991, Nature, 354, 43

\bibitem[Porth et al.(2014)]{porth14}
Porth, O., Komissarov, S. S., \& Keppens, R.
2014, MNRAS, 438, 278

\bibitem[Reach et al.(2006)]{reach06}
Reach, W. T., Rho, J., Tappe, A., et al.
2006, AJ, 131, 1479

\bibitem[Reynolds et al.(2009)]{reynolds09}
Reynolds, S. P., Borkowski, K. J., Green, D. A., 
et al.
2009, ApJ, 695, L149 

\bibitem[Reynolds \& Chevalier(1984)]{reynolds84}
Reynolds, S. P., \& Chevalier, R. A.
1984, ApJ, 278, 630

\bibitem[Reynolds et al.(2012)]{reynolds12}
Reynolds, S. P., Gaensler, B. M., \& Bocchino, F.
2012, SpSciRev, 166, 231

\bibitem[Reynolds \& Keohane(1999)]{reynolds99}
Reynolds, S. P., \& Keohane, J. W.
1999, ApJ, 525, 368

\bibitem[Reynolds et al.(1994)]{reynolds94}
Reynolds, S. P., Lyutikov, M., Blandford, R. D., \& Seward, F. D.
1994, MNRAS, 271, L1

\bibitem[Roberts et al.(2005)]{roberts05}
  Roberts, M. S. E., Lyutikov, M., Gaensler, B. M., et al.
  2005, in ``X-Ray and Radio Connections'' (eds. L.O. Sjouwerman and K.K. Dyer),
  published electronically by NRAO (http://www.aoc.nrao.edu/events/xraydio)

\bibitem[Roberts et al.(2004)]{roberts04}
  Roberts, M. S. E., Lyutikov, M., Kaspi, V. M., \& Tam, C. R.
  2004, BAAS, 36, 1522

\bibitem[Roberts et al.(2003)]{roberts03}
  Roberts, M. S. E., Tam, C. R., Kaspi, V. M., et al.
2003, ApJ, 588, 992

\bibitem[Rots(2009)]{rots09}
Rots, A.
2009, Determining the Astrometric Error in CSC Source Positions,
http://cxc.harvard.edu/csc/memos/files/Rots\_CSCAstrometricError.pdf

\bibitem[Salmon et al.(2014)]{salmon14}
    Salmon, J., Harmany, Z., Deladalle, C.-A., \& Willett, R.
    2014, J.~Math.~Imaging Vis., 48, 279

\bibitem[Sault \& Killeen(1999)]{sk99}
  Sault, R. J., \& Killeen, N. E. B.
  1999, The MIRIAD User's Guide (Sydney: ATNF)
    
\bibitem[Smith(2014)]{smith14}
Smith, N. 2014, ARA\&A, 52, 487

\bibitem[Smith et al.(2001)]{smith01}
  Smith, R. K., Brickhouse, N. S., Liedahl, D. A., \& Raymond, J. C.
  2001, ApJ, 556, L91

\bibitem[Stephenson \& Green(2002)]{stephenson02}
Stephenson, F. R., \& Green, D. A.
2002, Historical Supernovae and their Remnants (Oxford:
Oxford University Press)

\bibitem[Tam \& Roberts(2003)]{tam03}
Tam, C., \& Roberts, M. S. E. 
2003, ApJ, 598, L27

\bibitem[Tam et al.(2002)]{tam02}
Tam, C., Roberts, M. S. E., \& Kaspi, V. M.
2001, ApJ, 572, 202

\bibitem[Torii et al.(1997)]{torii97}
Torii, K., Tsunemi, H., Dotani, T., \& Mitsuda, K.
1997, ApJ, 489, L145

\bibitem[Torii et al.(1999)]{torii99}
  Torii, K., Tsunemi, H., Dotani, T., et al.
1999, ApJ, 523, L69

\bibitem[Vasisht et al.(1996)]{vasisht96}
Vasisht, G., Aoki, T., Dotani, T., Kulkarni, S. R., \& Nagase, F.
1996, ApJ, 456, L59 

\bibitem[Vink(2008)]{vink08}
  Vink, J.
  2008, ApJ, 689, 231

\bibitem[Wachter et al.(2010)]{wachter10}
  Wachter, S., Mauerhan, J. C., Van Dyk, S. D., et al.
  2010, AJ, 139, 2330
  
\bibitem[Weingartner \& Draine(2001)]{wg01}
Weingartner, J. C., \& Draine, B. T. 2001, ApJ, 548, 296

\bibitem[Williams et al.(2011)]{williams11}
Williams, B. J., Blair, W. P., Blondin, J. M., et al. 
2011, ApJ, 741:96

\end{thebibliography}
\end{document}